\documentclass{aa}
\usepackage{graphicx}
\usepackage{txfonts}

\usepackage{subcaption}
\usepackage{hyperref} 

\usepackage{placeins}

\hypersetup{colorlinks=true,citecolor=blue}

\begin{document}

\title{Constraints on the possible atmospheres on TRAPPIST-1 b: insights from 3D climate modeling}

\author{
    Alice Maurel\inst{1,2},
    Martin Turbet\inst{2,3},
    Elsa Ducrot\inst{4,5},
    Jérémy Leconte\inst{3},
    Guillaume Chaverot\inst{6,7},
    Gwenael Milcareck\inst{3},
    Alexandre Revol\inst{7,8},
    Benjamin Charnay\inst{5},
    Thomas J. Fauchez\inst{9,10,11}, \\
    Michaël Gillon\inst{12},
    Alexandre Mechineau\inst{3},
    Emeline Bolmont\inst{7,8},
    Ehouarn Millour\inst{2},
    Franck Selsis\inst{3},
    Jean-Philippe Beaulieu\inst{1},
    Pierre Drossart\inst{1}
}

\institute{
    \inst{1} Institut d’Astrophysique de Paris, Sorbonne Universit\'e, CNRS, 98 bis bd Arago, 75014 Paris, France \\
    \inst{2} Laboratoire de M\'et\'eorologie Dynamique/IPSL, CNRS, Sorbonne Universit\'e, Ecole Normale Sup\'erieure, Universit\'e PSL, Ecole Polytechnique, Institut Polytechnique de Paris, 75005 Paris, France \\
    \inst{3} Laboratoire d'astrophysique de Bordeaux, Univ. Bordeaux, CNRS, B18N, allée Geoffroy Saint-Hilaire, 33615 Pessac, France \\
    \inst{4} AIM, CEA, CNRS, Universit\'e Paris-Saclay, Universit\'e de Paris, F-91191 Gif-sur-Yvette, France \\
    \inst{5} LIRA, Observatoire de Paris, Universit{\'e} PSL, CNRS, Sorbonne Universit{\'e}, Univ. Paris Diderot, Sorbonne Paris Cit{\'e}, 5 place Jules Janssen, 92195 Meudon, France \\
    \inst{6} Univ. Grenoble Alpes, CNRS, IPAG, F-38000 Grenoble, France \\
    \inst{7} Centre pour la Vie dans l'Univers, Universit\'e de Gen\`eve, Geneva, Switzerland \\
    \inst{8} Observatoire de Gen\`eve, Universit\'e de Gen\`eve, Chemin Pegasi 51, 1290, Sauverny, Switzerland \\
    \inst{9} NASA Goddard Space Flight Center, 8800 Greenbelt Road, Greenbelt, MD 20771, USA \\
    \inst{10} Integrated Space Science and Technology Institute, Department of Physics, American University, Washington DC \\
    \inst{11} NASA GSFC Sellers Exoplanet Environments Collaboration \\
    \inst{12} Astrobiology Research Unit, Universit\'e de Li\`ege, All\'ee du 6 ao\^ut 19, Li\`ege, 4000, Belgium
}

\date{Submitted to A\&A}
\abstract{
\textit{Context.} JWST observations of the secondary eclipse of TRAPPIST-1 b at 12.8 and 15~$\mu$m revealed a very bright dayside. These measurements are consistent with an absence of atmosphere. Previous 1D atmospheric modeling also excludes -- at first sight -- CO$_2$-rich atmospheres. However, only a subset of the possible atmosphere types  has been explored, and  ruled out, to date.
Recently, a full thermal phase curve of the planet at 15~$\mu$m with JWST has also been observed, allowing for more information on the thermal structure of the planet.
\\
\textit{Aims.} We first looked for atmospheres capable of producing a dayside emission compatible with secondary eclipse observations. We then tried to determine which of these are compatible with the observed thermal phase curve.
\\
\textit{Methods.} We used a 1D radiative-convective model and a 3D global climate model (GCM) to simulate a wide range of atmospheric compositions and surface pressures. We then produced observables from these simulations and compared them to available emission observations.
\\
\textit{Results.} We found several families of atmospheres compatible at 2$\sigma$ with the eclipse observations. Among them, some feature a flat phase curve and can be ruled out with the observation, and some produce a phase curve still compatible with the data (i.e., thin N$_2$-CO$_2$ atmospheres, and CO$_2$ atmospheres rich in hazes). We also highlight different 3D effects that could not be predicted from 1D studies (redistribution efficiency, atmospheric collapse).
\\
\textit{Conclusions.} The available observations of TRAPPIST-1 b are consistent with an airless planet, which is the most likely scenario. A second possibility is a thin CO$_2$-poor residual atmosphere. However, our study shows that different atmospheric scenarios can result in a high eclipse depth at 15~$\mu$m. It may therefore be hazardous, in general, to conclude on the presence of an atmosphere from a single photometric point.}
  

\titlerunning{3-D Modeling constraints on the possible atmospheres of TRAPPIST-1 b}
\authorrunning{A. Maurel et al.} 

\maketitle

\section{Introduction}\label{sec:introduction}

Among the thousands of exoplanets discovered so far, only a few tens of them are temperate, with a size and/or a mass compatible with a rocky composition. Their small size makes them difficult to detect and even more difficult to characterize. \cite{Hu:2024} report the potential detection of a secondary atmosphere on the hot planet 55 Cancri e. A sulfur-rich atmosphere may have been detected on L 98-59 d \citep{Banerjee:2024,Gressier:2024}; however,   no definitive detection of a secondary atmosphere has been made to date on a temperate Earth-sized rocky exoplanet. The James Webb Space Telescope (JWST) has the requisite sensitivity to probe the atmospheres of a few specific temperate rocky planets, mostly through transit spectroscopy and occultations (i.e., secondary eclipses). Those specific targets are exoplanets orbiting the coolest and smallest close-by stars, which are called red dwarf stars. Red dwarf stars are favorable targets \citep{Triaud:2021,Nutzman_charbonneau:2008,Sullivan:2015,Barclay:2018}, because (1) their small size makes it easier to detect and characterize small transiting planets, (2) their lower luminosities result in more frequent planetary transits and occultations for the same stellar irradiation, (3) they are the most common objects in the Galaxy, and (4) temperate planets around them are more likely to be short-period rocky planets, allowing us to accumulate many transits and occultations or to perform phase curve observations. 

With its seven transiting temperate rocky planets around the coolest host star known to date, the TRAPPIST-1 system is the most studied system of this kind \citep{Turbet:2020review}. Thanks to very precise measurements of radii \citep{Ducrot:2020} and masses \citep{Agol2021}, TRAPPIST-1 planets are the most well-known rocky planets outside  our Solar System. Since its discovery in 2016-2017 \citep{Gillon:2016,Gillon:2017}, a large number of studies has assessed the possibility of atmospheres on these planets \citep{Turbet:2020review}, and their detectability with the JWST \citep{Lustig-Yaeger:2019,Fauchez:2019}. The host star, with its high flux in the extreme ultraviolet and X-rays (XUV) and its long and hot pre-main sequence phase, is expected to result in significant atmospheric escape \citep{Bolmont2017, Bourrier:2017}. However, depending on the formation and evolution pathways of the planets, they could have retained a part of their primary atmosphere or formed a secondary atmosphere (see \cite{Moore:2020, Kite:2020} for the evolution of rocky planets and potential secondary atmospheres, \cite{Coleman:2019,Turbet:2020review} for the particular case of TRAPPIST-1).

Considering their transiting nature combined with the infrared brightness (K = 10.3) and the Jupiter-like size of their host star ($\sim$~12\% the radius of Sun), they have been designated as the most promising candidates for the first detailed study of rocky Earth-sized exoplanets with JWST \citep{Gillon:2020}. Since its launch, all the TRAPPIST-1 planets have been observed in transit (JWST programs GTO 1201, GTO 1331, GO 1981, GO 2420, GO 2589, GO 6456). Unfortunately, the activity of the star can mimic the signatures from the planet's atmosphere in transit \citep{Lim:2023}. The host star's surface presents inhomogeneities that can cause artificial spectral features in the transit, due to unocculted spots on the star rather than an atmosphere on the planet \citep{Rackham:2018}. One way out of this problem is to observe the planets in emission rather than in transmission spectroscopy. In this case the surface inhomogeneities of the star have a much weaker impact, although not inexistent \citep{Fauchez:2025}. The planets in the TRAPPIST-1 system that are hot enough to be observed with this method by the JWST are TRAPPIST-1 b and c. The two planets' emissions have been observed through their secondary eclipses  in the 15~$\mu$m MIRI filter (JWST programs GTO 1177 and GO 2304) and in the 12.8~$\mu$m filter (GTO 1279 and GO 5191). 

The 15~$\mu$m eclipse observations of TRAPPIST-1 b  already ruled out --at first sight-- a thick CO$_2$ atmosphere \citep{Greene:2023}, but they still left space for thinner atmospheres with different compositions \citep{Ih:2023}. Taking into account the second photometric point, at 12.8~$\mu$m, \cite{Ducrot:2024} pointed out that secondary eclipses can also theoretically be compatible with a CO$_2$-dominated atmosphere rich in hazes. A way to break this degeneracy inherent to dayside emission is to observe phase curves in order to get the nightside thermal emission of the planet \citep{Hammond:2025}.

\begin{figure}
    \centering
    \includegraphics[width=\linewidth]{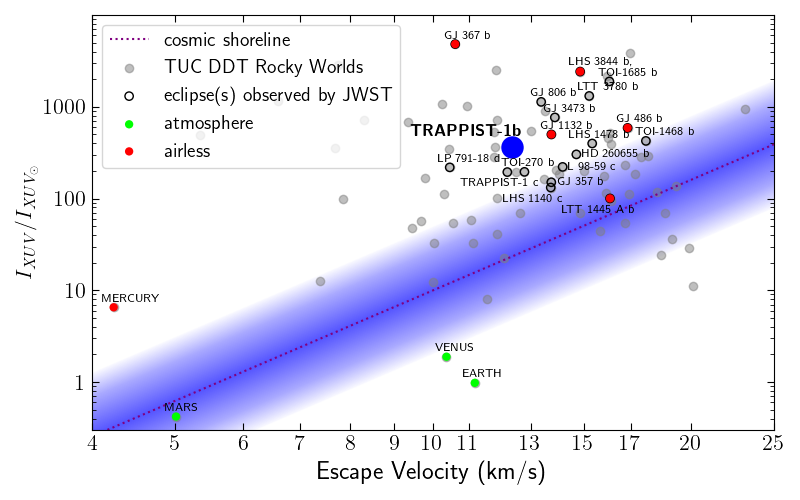}
    \caption{Exoplanets under consideration for the Rocky Worlds DDT. Planets above the cosmic shoreline are expected to have lost their atmosphere. The circled exoplanets have already been observed in emission during JWST's first three cycles. The exoplanets for which the observations concluded there is no atmosphere are in red. We also added the Solar System terrestrial planets. The XUV irradiation is computed from \cite{Zahnle:2017}.}
    \label{fig:cosmic_shoreline}
\end{figure}

More recently, a phase curve of the system has been observed and carefully selected to be centered on the eclipses of TRAPPIST-1b and c (GO 3077).
In this work, we present 1D radiative-convective and 3D numerical climate simulations of various types of atmosphere of TRAPPIST-1 b and compare them to all the available thermal emission observations (eclipses and phase curve). Although this work is an in-depth study of a single planet (the only moderately irradiated rocky planet for which we have such a large amount of observation data), its approach and results are meant to be broadly applicable to other temperate rocky exoplanets for which we have emission observations. This includes TRAPPIST-1c, but also LHS-1140 c, LHS-1478 b, GJ-3473 b, GJ-357 b, HD-260655 b, L-98-59 c, LTT-3780 b, TOI 1468 b, and TOI-270 b (the nine planets of the Hot Rocks Survey, GO 3730); LP 791-18 d (GO 6457); and Gl 486 b (GO 1743). Moreover, a 500-hour Director's Discretionary Time program on JWST (Rocky Worlds DDT) will provide more mid-infrared observations of secondary eclipses of small rocky exoplanets, in order to probe their ability to maintain an atmosphere(Fig.~\ref{fig:cosmic_shoreline}).

\section{Methods}\label{sec:methods}

Starting from the secondary eclipse observations at 15 and 12.8~$\mu$m \citep{Greene:2023,Ducrot:2024}, we attempted to identify all the possible families of atmosphere of TRAPPIST-1 b that could account for the high dayside emission flux measured in secondary eclipses. For this, we used a suite of numerical climate models, ranging from (1) 1D radiative-convective simulations designed for a fast exploration of the parameter space to (2) 3D global climate calculations to more reliably simulate the thermal structure and temperature contrasts across the planet, which play a major role in the planet's emission signal. These simulations were then post-processed with radiative transfer tools to generate synthetic eclipse spectra and thermal phase curves, for comparison to the observations.

In practice, we first carried out dozens of numerical climate simulations with the 1D model \texttt{exo\_k} (see Section~\ref{sec:methods/1Dmodel}) for a wide range of atmospheric compositions, and computed secondary eclipse spectra (see Section~\ref{sec:methods/postprocess}) for a large range of plausible atmospheric redistribution factors. For atmospheric compositions that yield good match with eclipse data, we used a 3D climate model to produce much more realistic simulations (see Section~\ref{sec:methods/gPCM}) with self-consistent heat redistribution, verified or not the good match with the eclipse data, and calculated thermal phase curves in the 15~$\mu$m MIRI filter. The list of all the 3D simulations used in this work is given in Table~\ref{table:method_simulations_list}.

    \begin{table*}
\caption{Summary of all the GCM simulations performed in this work.}
\centering
\begin{tabular}{lcccr}
   \hline
   Composition & Surface Pressure (bar) & $\chi^2$ eclipse & $\chi^2$ phase curve & $\chi^2$ joint\\
   \hline
   Pure CO$_2$                      & 1             & 23(6.5$\sigma$)    & 43(11$\sigma$)     & 35(12$\sigma$) \\
   Pure CO$_2$                      & 0.1*          & 17(5.6$\sigma$)    & 38(10$\sigma$)     & 30(11$\sigma$) \\
   Pure CO$_2$                      & 0.1**& 17(5.6$\sigma$)    & 38(10$\sigma$)     & 30(11$\sigma$) \\
   Pure CO$_2$                      & 0.1**& 17(5.6$\sigma$)    & 38(10$\sigma$)     & 30(11$\sigma$) \\
   Pure CO$_2$                      & 0.01*         & 10(4.1$\sigma$)    & 25(8.2$\sigma$)    & 19(9$\sigma$) \\
   N$_2$ + 1\% CO$_2$               & 1             & 25(6.7$\sigma$)    & 50(12$\sigma$)     & 40(13$\sigma$) \\
   N$_2$ + 1\% CO$_2$               & 0.1           & 15(5.2$\sigma$)    & 37(10$\sigma$)     & 29(11$\sigma$) \\
   N$_2$ + 1\% CO$_2$               & 0.01          & 3.7(2.2$\sigma$)   & 5.5(3.3$\sigma$)   & 4.8(3.7$\sigma$) \\
   N$_2$ + 100 ppm CO$_2$           & 1             & 14(4.8$\sigma$)    & 37(10$\sigma$)     & 27(11$\sigma$) \\
   N$_2$ + 100 ppm CO$_2$           & 0.1           & 3.3(2.1$\sigma$)   & 5.1(3.2$\sigma$)   & 4.4(3.5$\sigma$) \\
   N$_2$ + 100 ppm CO$_2$           & 0.01*         & 3.6(2.2$\sigma$)   & 0.44(<1$\sigma$) & 1.7(1.5$\sigma$) \\
   N$_2$ + 1 ppm CO$_2$             & 1             & 2.4(1.7$\sigma$)   & 4.5(2.9$\sigma$)   & 3.7(3$\sigma$) \\
   N$_2$ + 1 ppm CO$_2$             & 0.1           & 3.7(2.3$\sigma$)   & 0.17(<1$\sigma$)  & 1.6(1.4$\sigma$) \\
   N$_2$ + 1 ppm CO$_2$             & 0.01          & 4(2.3$\sigma$)     & 0.62(<1$\sigma$) & 2(1.7$\sigma$) \\
   N$_2$+0.1ppm CO$_2$              & 1             & 2.8(1.9$\sigma$)   & 0.72(<1$\sigma$) & 1.6(1.4$\sigma$) \\
   N$_2$+0.01ppm CO$_2$             & 10            & 2.2(1.6$\sigma$)   & 17(6.5$\sigma$)    & 11(6.3$\sigma$) \\
   N$_2$+20\%CH$_4$+0.2\%C$_2$H$_4$ & 10            & 1.8(1.4$\sigma$)   & 34(9.7$\sigma$)    & 21(9.5$\sigma$) \\
   N$_2$+40\%CH$_4$+0.4\%CO$_2$     & 1             & 1.2(1$\sigma$)     & 20(7.2$\sigma$)    & 12(6.9$\sigma$) \\
   CO$_2$+Hazes (Model 1)           & 1             & 0.35(<1$\sigma$) & 0.44(<1$\sigma$) & 0.4(<1$\sigma$) \\
   CO$_2$+Hazes (Model 2)           & 1             & 1.8(1.4$\sigma$)   & 4.7(3$\sigma$)     & 3.5(2.9$\sigma$) \\
   CO$_2$+Dust                      & 1             & 1.7(1.3$\sigma$)   & 1.1(<1$\sigma$)  & 1.3(1.2$\sigma$) \\
   CO$_2$+Tholins                   & 1              & 15(5.1$\sigma$)   & 29(8.8$\sigma$)    & 23(10$\sigma$) \\
   \hline
   Airless, a=0.2                   & --            & 1.7(1.3$\sigma$)   & 1.5(1.3$\sigma$)   & 1.6(1.4$\sigma$) \\
   \hline
\end{tabular}
    \tablefoot{Surface pressures flagged with * correspond to collapsing atmospheres and flagged with  ** corresponds to simulations with a circulation of the substellar point. We added the computed reduced $\chi^2$ of the simulated atmospheres (see Section \ref{sec:method/comp_data}) and corresponding standard deviation in parentheses, computed as in \cite{ducrotgillon:2025}}.
    \label{table:method_simulations_list}
\end{table*}

The models and tools used in our study are detailed in the subsections below.

\subsection{3D simulations with the generic PCM}\label{sec:methods/gPCM}

For the 3D atmospheric simulations of TRAPPIST-1 b, we used the \texttt{Generic Planetary Climate Model} (\texttt{Generic PCM}), a 3D numerical climate model which has now been extensively applied to a large diversity of exoplanets \citep{Wordsworth:2011ajl,Charnay:2015a,Bolmont:2016aa, Turbet:2018,Charnay:2021,Turbet:2023aa,Teinturier:2024}.

\subsubsection{Model parameters}
For the stellar, planetary and orbital parameters, we used the results of \cite{Agol2021}. We assumed (unless otherwise stated) a circular, 0$^{\circ}$ obliquity, tidally locked orbit. Table~\ref{table:method_param} summarizes all the parameters used to set up our simulations. Sensitivity tests for some of these parameters are detailed in Section \ref{sec:results}.

\subsubsection{Model resolutions}

For all our simulations, we used a standard horizontal resolution of $72~\times~46$ in longitude~$\times$~latitude. For the vertical resolution, our standard simulations use 40 pressure layers, but for some simulations we had to change the number and distribution of pressure layers (suppressing layers for the simulation with atmospheric collapse~; adding and refining the upper layers grid for the simulation with a thermal inversion). Given that we simulated a broad range of atmospheric compositions in the \texttt{Generic PCM}, with various radiative and dynamical timescales involved, we adapted the time step (ranging from $\sim$0.5s to $\sim$60s) for all the simulations.

\begin{table*}
\caption{Summary of the parameters used for GCM simulations of TRAPPIST-1 b performed in this study.}
\centering
\begin{tabular}{lr}
   \hline
   Physical parameters & Values \\
   \hline
   Mass (M$_{\oplus}$) & 1.374 \\
   Radius (M$_{\oplus}$) & 1.116 \\
   Insolation (S$_{\oplus}$) & 4.153 \\
   \hline
   Rotation mode & synchronous/substellar circulation\\
   Rotation rate (rad~s$^{-1}$) & $4.8128 \times 10^{-5}$\\
   Obliquity ($^\circ$) & 0 \\
   Orbital eccentricity & 0  \\
   \hline
   Bare ground albedo & 0.1 \\
   Ground thermal inertia (J~m$^{-2}$~s$^{-1/2}$~K$^{-1}$) & 250 \\
   Surface Topography & flat \\
   Surface roughness coefficient (m) & 0.01 \\
   Internal Heat Flux (W~m$^{-2}$) & 0 \\
   \hline
   Stellar effective temperature (K) & 2566 \\
   Stellar spectrum &  2600 K BT-Settl with Fe/H = 0\\
   \hline
\end{tabular}

\label{table:method_param}
\end{table*}

\subsubsection{Radiative transfer}

Regarding the radiative transfer calculations, the model is based on the correlated-k approach \citep{Fu:1992}. For the correlated-k tables, we use 31 spectral bands in the visible-near infrared (0.3 to 6.6~$\mu$m), corresponding to the stellar incoming radiation, and 40 in the thermal infrared (1 to 50~$\mu$m), corresponding to the planet emission wavelengths. This corresponds to a spectral resolution of R=10 for both regimes. We used 16 gauss points for the cumulated distribution function (g) integration, 20 for the N$_2$-CO$_2$ atmospheres. 

The correlated-k tables used are summarized in Table~\ref{table:method_opacites}. 
The correlated-k tables of N$_2$+CH$_4$+C$_2$H$_4$ and N$_2$+CH$_4$+CO$_2$ have been calculated by combining single gas opacities from Exomol \citep{exomol-CH4,exomol-C2H4,exomol-CH4,exomol-CO2,exomol-CO2_1} using \texttt{exo\_k}. For N$_2$+CO$_2$ mixtures, original correlated-k tables have been calculated from line-by-line calculations of the high-resolution absorption spectra of CO$_2$ broaden by N$_2$ (Chaverot et al. in prep.). Absorptions have been computed following a Voigt profile with empirical corrections in the far wings \citep[e.g.,][]{tran_measurements_2018} based on laboratory measurements \citep{burch_absorption_1969,perrin_temperature-dependent_1989,tran_measurements_2011}. For all atmospheric compositions, we also included the effect of collision-induced absorptions (CIAs) of CO$_2$-CO$_2$ \citep{Gruszka:1997,Tran:2024} N$_2$-N$_2$ \citep{Karman:2019}, CH$_4$-CH$_4$ \citep{Karman:2019}, CO$_2$-CH$_4$ \citep{Turbet:2020spectro} and CH$_4$-N$_2$ \citep{Karman:2019}.

\begin{table*}
\caption{Summary of the opacity cross sections and continuum look-up tables used for the various molecules in the atmospheric simulations performed in this study.}
\centering
\begin{tabular}{lp{14cm}}
   \hline
   Composition  & References \\
   \hline
   k-correlated tables &  \\
   \hline
   Pure CO$_2$ & This work \\
   N$_2$+CO$_2$ & This work, excepting 0.1ppm of CO$_2$ from Exomol \citep{exomol-CO2,exomol-CO2_1}  \\
    N$_2$+CH$_4$+CO$_2$ & Exomol \citep{exomol-CH4,exomol-CO2,exomol-CO2_1}  \\
    N$_2$+CH$_4$+C$_2$H$_4$ & Exomol \citep{exomol-CH4,exomol-C2H4} \\
   CO$_2$+Hazes (simplified) & This work \\
   CO$_2$+Hazes (Tholins) & This work + \cite{Khare:1984,Drant:2024}\\
   CO$_2$+Dust & This work + \cite{Forget:1999}\\
   CO$_2$+H$_2$SO$_4$ & This work + \cite{Myhre:2003}\\

   \hline
   CIA & \\
   \hline
   CO$_2$-CO$_2$ & \cite{Gruszka:1997,Tran:2024} \\
   N$_2$-N$_2$ & \cite{Karman:2019}\\
   CH$_4$-CH$_4$ & \cite{Karman:2019}\\
   CH$_4$-N$_2$ & \cite{Karman:2019} \\
   CO$_2$-CH$_4$ & \cite{Turbet:2020spectro} \\

   \hline
\end{tabular}
\label{table:method_opacites}
\end{table*}

\subsubsection{Convection and cloud formation}
The subgrid-scale dynamical processes (dry turbulent mixing and convection) are taken into account through a convective adjustment scheme that adjusts unstable temperature profile to the dry adiabat \citep{Forget:1999}.
For the study of the collapse of the CO$_2$ atmospheres, we activated the CO$_2$ condensation scheme \citep{Forget:2013}. This scheme is performed when a grid cell reaches 100\% of saturation. CO$_2$ when condensing can form CO$_2$-ice clouds and/or surface frost. Cloud ice particles can sediment and accumulate at the surface, and the surface ice layer can itself re-sublimate in the atmosphere, depending on the thermodynamic conditions. The CO$_2$ ice albedo at the surface was set to 0.5 \citep{Forget:2013}, although this value has a limited impact given that CO$_2$ ice forms in the nightside, where there is no starlight to reflect. The CO$_2$ aerosols thus formed were radiatively active, and the number mixing ratio of cloud condensation nuclei fixed to 10$^5$~kg$^{-1}$, but tested the sensitivity by running simulations at 10$^2$~kg$^{-1}$ and 10$^8$~kg$^{-1}$, and also by doing a test in which we turned off the radiative effect of CO$_2$ ice clouds.

\subsubsection{Hazes and aerosols}\label{sec:methods/gPCM/hazes}

For some simulations, we implemented and tested the effect of high-altitude aerosols. We first explored the effect of idealized hazes \citep{Ducrot:2024}. For this, we added a haze term to the correlated-k table of a pure CO$_2$ atmosphere, of the shape 
\begin{equation}
    \sigma=f_\mathrm{haze}~\kappa~(\lambda_0/\lambda)^2~(M_\mathrm{CO_2}/\mathcal{N}_a)
    \label{eq:haze_simple}
\end{equation}
with $\kappa=0.5~\mathrm{cm}^2/g$ and $\lambda_0=1~\mu \mathrm{m}$ \citep{Ducrot:2024}. The haze vertical mixing ratio profile was assumed to be constant. We varied different factors (haze mixing ratio profile, haze factor $f_\mathrm{haze}$, see Equation \ref{eq:haze_simple}) in 1D (Appendix~\ref{sec:app/hazes/parameters}).

We also explored more realistic optical properties for different hazes and aerosols (Table~\ref{table:method_opacites}). In this case the radiative transfer code computes the aerosols' contribution from their absorption coefficient, single scattering albedo and asymmetry factor. These parameters depend on the radius distribution of the particles. For both \texttt{exo\_k} and the \texttt{Generic PCM}, or all aerosols, we prescribed an upper and lower pressure layer, and the aerosols follow the atmospheric scale height between them. The parameters (upper and lower pressure layers, volume mixing ratio, and size distribution of aerosols) are detailed is the dedicated sections (Section \ref{sec:results/inversion/comp}, Appendix \ref{sec:app/hazes/parameters}).

\subsection{1D radiative-convective simulations with \texttt{exo\_k}}\label{sec:methods/1Dmodel}

We used the \texttt{exo\_k} python library \citep{Leconte:2021} that includes a 1D atmospheric evolution model \citep{Selsis:2023}, in order to explore a wide range of possibilities in a limited amount of time.
We performed $\sim$~80 simulations in which we varied atmospheric composition (different proportions of N$_2$, CH$_4$, CO$_2$, C$_2$H$_4$, C$_2$H$_6$, NH$_3$, SO$_2$) and total atmospheric pressure (from 0.1 to 10~bar).
We used the same stellar, planetary and orbital parameters as described above for the \texttt{Generic PCM} (Table~\ref{table:method_param}). We also used the same correlated-k tables as in the \texttt{Generic PCM} (Table~\ref{table:method_opacites}), also with a spectral resolution of 10, ranging between 0.3 and 50~$\mu$m). We only tested dry atmosphere cases.

We parameterized the 3D redistribution for the incoming stellar flux with a multiplication factor $f$, that spans from 2/3 (no redistribution) to 1/4 (full redistribution):
$$ISR=f \times \sigma T_s ^4 \left( R_s/a\right)^2$$

\subsection{Post-processing tools}\label{sec:methods/postprocess}

We computed the emission spectra of the modeled 1D atmospheres with \texttt{exo\_k}, and for the 3D atmospheres we used \texttt{Pytmosph3R} \citep{Falco:2022}, a python library based on \texttt{exo\_k} that computes observables (eclipse spectra, phase curves) directly from the \texttt{Generic PCM} 3D outputs.
When used in emission or phase curve mode, \texttt{Pytmosph3R} computes the emission for each column of the \texttt{Generic PCM} output, weighted by the angle of the observer.
We used a stellar spectrum from the SPHINX model grid \citep{Iyer:2023} interpolated to TRAPPIST-1's parameters. This spectrum, different from the one used in the GCM, was chosen for his closeness to the measured flux at 15~$\mu$m, allowing for a more accurate planet-to-star flux ratio - a criterion that is not sensitive in the GCM simulations, the stellar flux in the thermal infrared having a negligible effect on the GCM results.

We used correlated-k tables from the same opacity data (Table~\ref{table:method_opacites}) at higher spectral resolution ($R~\sim~100$) for the computation of the observables. To compare with the observations, we integrated the flux in the MIRI observed bands (F1280W and F1500W) with the available JWST MIRI filters documentation. We also computed the planet-to-star flux ratio by dividing the computed planet flux by the stellar flux. For this, we normalized the stellar spectrum with the 15~$\mu$m MIRI measurements of the stellar flux. At 12.8~$\mu$m, we find that the computed stellar flux is slightly lower than the measured flux (<2\%).

\subsection{Comparison to data}\label{sec:method/comp_data}
We compared our synthetic observables to the available JWST emission observations. First, to the eclipse depths of TRAPPIST-1 b at 15~$\mu$m \citep{Greene:2023} and 12.8~$\mu$m \citep{Ducrot:2024} (Fig.~\ref{fig:obs_families}); second, to the 15~$\mu$m thermal phase curve of b extracted from the observed combined phase curve of TRAPPIST-1 b and c \citep{ducrotgillon:2025}. \cite{ducrotgillon:2025} performed four independent analyses, each one divided in several subanalyses with different assumptions. We chose to compare our results to analysis \#1-MG because the analysis of MG was highlighted in the paper and only the subanalysis \#1 allowed for an atmosphere on TRAPPIST-1 b. We note that contrary to \cite{Ducrot:2024}, our best-fit airless albedo is not 0.2, but $\sim 0.4$ (see Section \ref{sec:results/co2/obs/eclipse}).
For the phase curve data, shown in Fig.~\ref{fig:phc_data}, we chose to compare three parameters extracted from the phase curve: the maximal flux, the minimal flux, and the peak offset to evaluate the match between our simulations and the observation. The extracted phase curve of the planet b is quite dependent on the analysis pipeline -- especially in this case where one must extract two signals (planet b and c) from one phase curve. Furthermore, any extracted phase curve is only one of the thousands of outputs of the MCMC analysis. On the other hand, the three variables extracted correspond to the posterior probability distribution functions of the analysis, and are thus much more robust. These quantities are also physically interpretable (amplitude corresponds to heat redistribution, offset to shift of the hottest hemisphere).
We note that due to the complexity of the GCM simulations, we cannot perform a bayesian retrieval like with a simpler, 1D model. We simply tried to assess the match of our different simulations with the observations.
For this, we computed the $\chi ^{2}$ as
\begin{equation}
    { \chi ^{2}=\sum _{i}{\frac {(O_{i}-M_{i})^{2}}{\sigma _{i}^{2}}}}
    \label{eq:chi2} ,
\end{equation}

where $O_i$ are the data points of the observations, $\sigma_i$ their 1$\sigma$ uncertainty, and $M_i$ the models results.

For the eclipse, the two data points correspond to the eclipse depths measured at 12.8 and 15~$\mu$m (\citep{Ducrot:2024}.
For the phase curve, we extracted the dayside flux, nightside flux and peak offset from the analysis \#1- MG of \cite{ducrotgillon:2025} and compared each of them.

We note that this $\chi^{2}$ is a metric aiming at comparing the match of the different simulated atmospheres with the data, relatively to each other. We cannot provide an absolute quantification of the mismatch that could be compared with other papers, because our approach is different from 1D models, and we are not performing a proper fit, but rather evaluating a few realizations that we wanted to explore.

\section{Results}\label{sec:results}
\begin{figure*}
     \centering
     \begin{subfigure}[b]{0.49\textwidth}
         \centering
         \includegraphics[width=\textwidth]{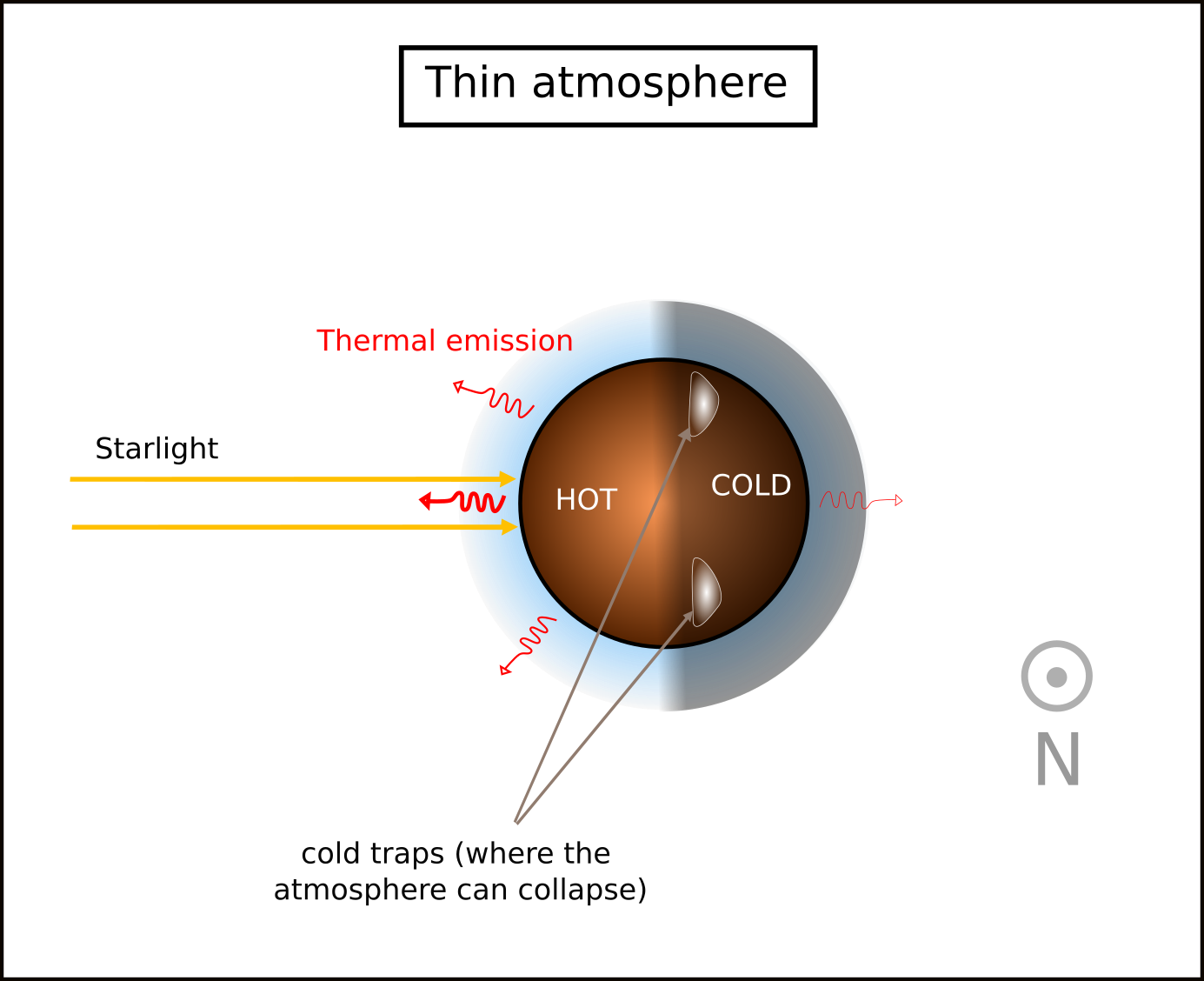}
         \caption{Case 1a: Thin residual atmospheres}
     \end{subfigure}
     \hfill
          \begin{subfigure}[b]{0.49\textwidth}
         \centering
         \includegraphics[width=\textwidth]{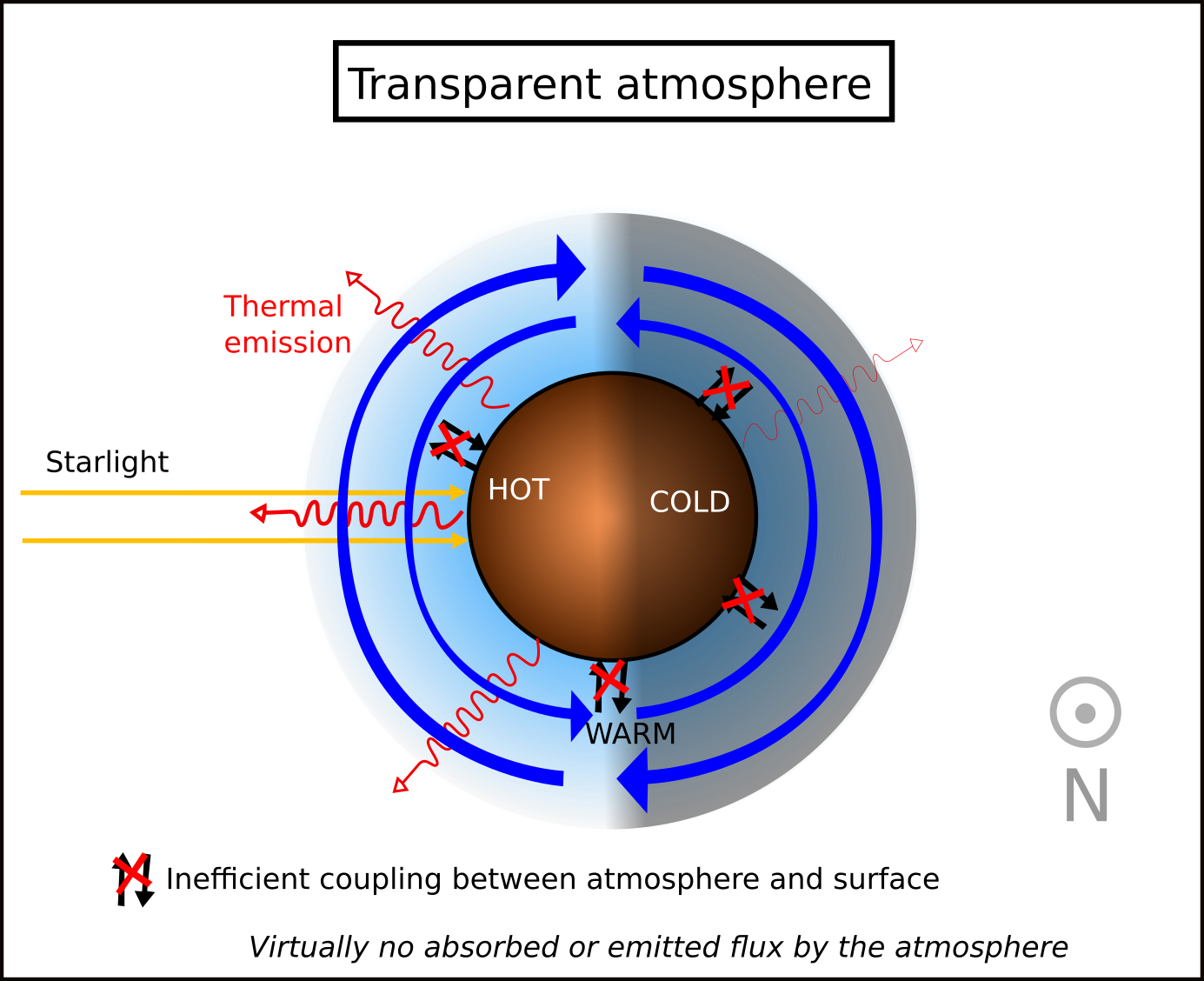}
         \caption{Case 1b: Non-condensible thick transparent atmospheres}
     \end{subfigure}
     \hfill
     \begin{subfigure}[b]{0.49\textwidth}
         \centering
         \includegraphics[width=\textwidth]{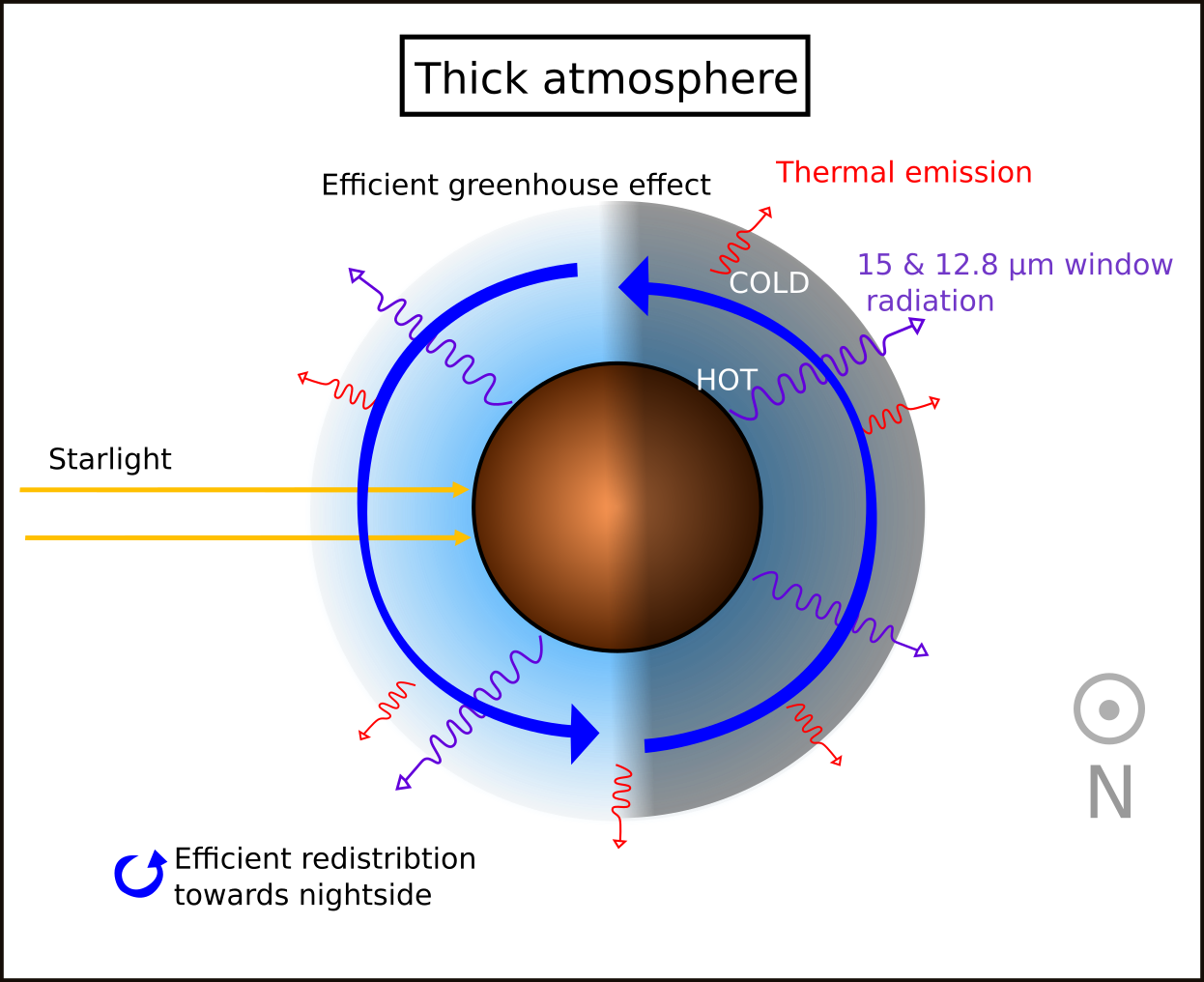}
         \caption{Case 2: Thick, greenhouse efficient atmospheres}
     \end{subfigure}
     \hfill
     \begin{subfigure}[b]{0.49\textwidth}
         \centering
         \includegraphics[width=\textwidth]{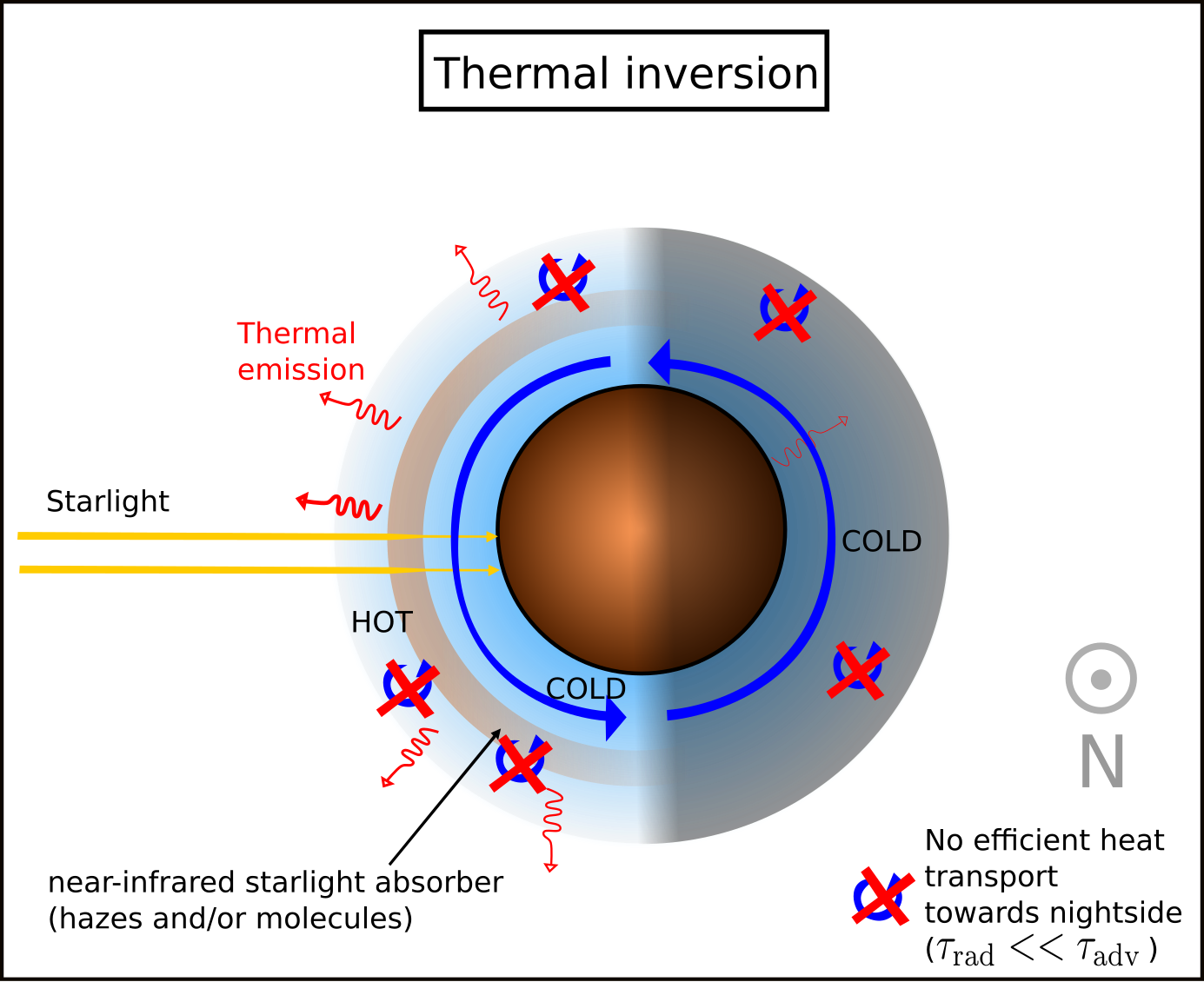}
         \caption{Case 3: Atmospheres with a strong thermal inversion}
     \end{subfigure}
        \caption{Sketches of the different atmospheric scenarios studied in depth in this work, as seen from the North pole.}
        \label{fig:schemas_atmos}
\end{figure*}

We identify four distinct families of atmospheres that can produce a strong dayside emission matching the secondary eclipse MIRI observations \citep{Greene:2023,Ducrot:2024}. These four families are summarized in the sketches of Fig.~\ref{fig:schemas_atmos} and listed below:
\begin{itemize}
    \item Thin, residual atmospheres: the thin atmosphere produces very little heat redistribution and greenhouse effect. In this case, the infrared brightness temperature is very close to the surface temperature, where most of the emission comes from. Here the planet behaves very closely to a bare rock planet.
    \item Non-condensible, thick, transparent atmospheres: the observer sees through the atmosphere directly the surface emission of the planet. In the absence of condensation and radiatively active species, the surface and the atmosphere only interact through sensible heat exchange, which is limited. Thus, the atmosphere can produce very efficient heat redistribution but does not interact sufficiently with the surface to deviate the surface temperature field from the bare-rock case. Such transparent atmospheres are difficult to probe remotely, except perhaps through Rayleigh scattering at visible wavelengths.    
    \item Thick, greenhouse efficient atmospheres: the atmosphere is thick, and produces a very efficient greenhouse effect. The large scale dynamics of such thick atmospheres produces efficient horizontal mixing, resulting in very efficient heat redistribution. As a result, the day-to-night temperature contrast is very low. In such thick atmospheres, the temperature of the lower atmosphere can far exceed the brightness temperature measured by MIRI, due to the powerful greenhouse effect of the atmosphere. For such atmospheres to fit the eclipse data, the molecules present in the atmosphere must have absorption windows that let through the emission from the deeper, warmer layers of the atmosphere.
    \item Atmospheres with a strong thermal inversion: the atmosphere possesses species (e.g., CH$_4$ or high-altitude hazes) that absorb efficiently at the wavelengths of the emission of the host star. A significant part of the stellar radiation is absorbed high in the atmosphere, producing strong shortwave heating, causing a thermal inversion. In this case, the CO$_2$ present in the atmosphere, with a strong absorption band at 15~$\mu$m, can now produce a strong signature in emission.
\end{itemize}

\begin{figure}
    \centering
    \includegraphics[width=\linewidth]{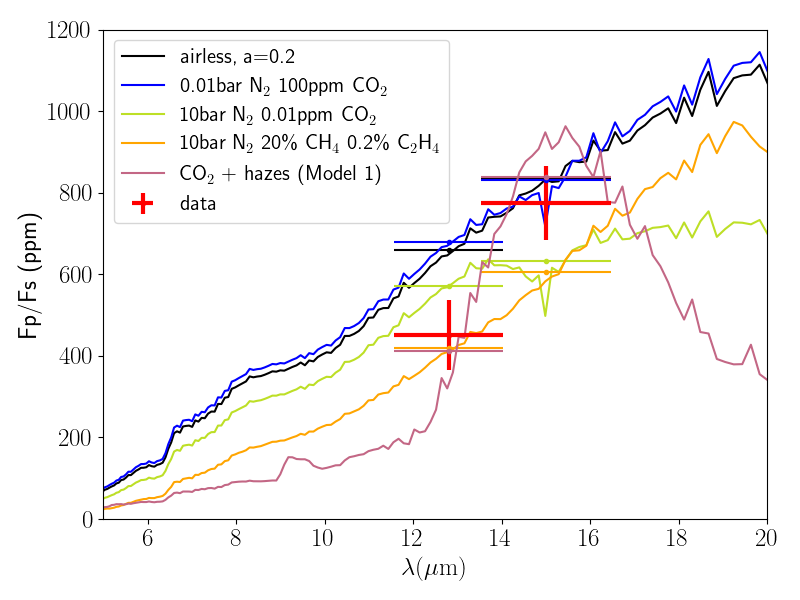}
    \caption{Dayside emission vs wavelength for a bare rock and for a representative of each family studied in this paper. The red data points and associated 1$\sigma$ error bars are from \cite{Ducrot:2024}.}
    \label{fig:obs_families}
\end{figure}

\begin{figure}
    \centering
    \includegraphics[width=\linewidth]{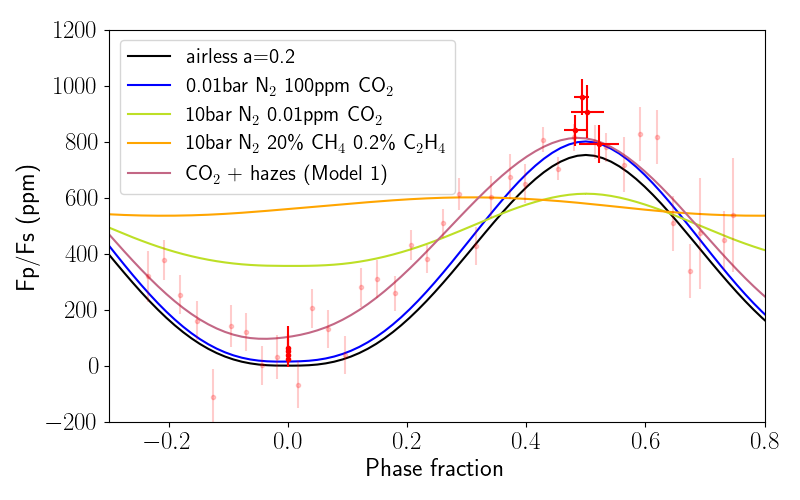}
    \caption{Phase curves of the simulations from Fig.~\ref{fig:obs_families} and data from \cite{ducrotgillon:2025}. The points in transparency correspond to the binned results of one  of the MCMC outputs of analysis (\#1- MG) of the paper. The non-transparent points are the maximum flux, minimum flux, and peak offset with 1$\sigma$ uncertainties resulting from different analyses of the paper.}
    \label{fig:phc_data}
\end{figure}

Fig.~\ref{fig:obs_families} shows the synthetic eclipse depths for four arbitrarily selected cases from the four families of atmospheric scenarios as a proof of concept. All of them can match the eclipse depth observations at the 2$\sigma$ confidence level.

The following subsections describe each of these atmospheric scenarios, show why they match the eclipse depth measurements, and how phase curves can be used to discriminate between them.

\subsection{Residual thin and thick transparent atmospheres}\label{sec:results/co2}
\subsubsection{Principle}
The first two families of atmospheres that can match the measured eclipse depths are residual, thin atmospheres and thick, transparent atmospheres. To explore these scenarios quantitatively, we have decided to focus on CO$_2$-N$_2$-dominated atmospheres. We chose this composition for three distinct reasons: (1) first, they were initially proposed as good candidate molecules for a putative TRAPPIST-1 b's atmosphere \citep{Turbet:2020review,Krissansen-Totton:2022}~; (2) second, these are compositions that have been explored in detail in previous theoretical studies of TRAPPIST-1 b \citep{Ih:2023}; and (3) third, because CO$_2$ is the strongest absorber at 15~$\mu$m and a condensable species, which is the most conservative case for our reasoning and makes it a good representative of this category.
The principle and mechanisms at stake that we describe here can be transposed to other compositions of residual and/or transparent atmospheres (see Section \ref{sec:discussion}).

To do this, we created a grid of 3D GCM simulations of TRAPPIST-1 b for a wide range of CO$_2$ and N$_2$ partial pressures. Our results are described in the following subsections.

\subsubsection{A grid of N$_2$+CO$_2$ atmospheres}\label{sec:results/co2/n2-co2}

We performed simulations varying total pressure (0.01 to 1~bar) and CO$_2$ volume mixing ratio (VMR) (1 ppm to 100\%), using a grid inspired from \cite{Ih:2023}. For each 3D simulation, we computed the eclipse depth spectrum, and the phase curve in the 15~$\mu$m (F1500W) MIRI filter. Compared to \cite{Ih:2023}, we added two extra points to specifically study the case of thick, ``transparent'' atmospheres with a low, constant CO$_2$ partial pressure of $10^{-7}$~bar, for a total pressure of 1 and 10~bar A specific study of these cases is presented in Appendix~\ref{sec:app-transparent_atmos}. Fig.~\ref{fig:N2-CO2-grid} shows the grid of N$_2$+CO$_2$ atmosphere simulations we explored, and summarizes the various observational and theoretical constraints discussed in the following subsections.

\begin{figure}
    \centering
    \includegraphics[width=\linewidth]{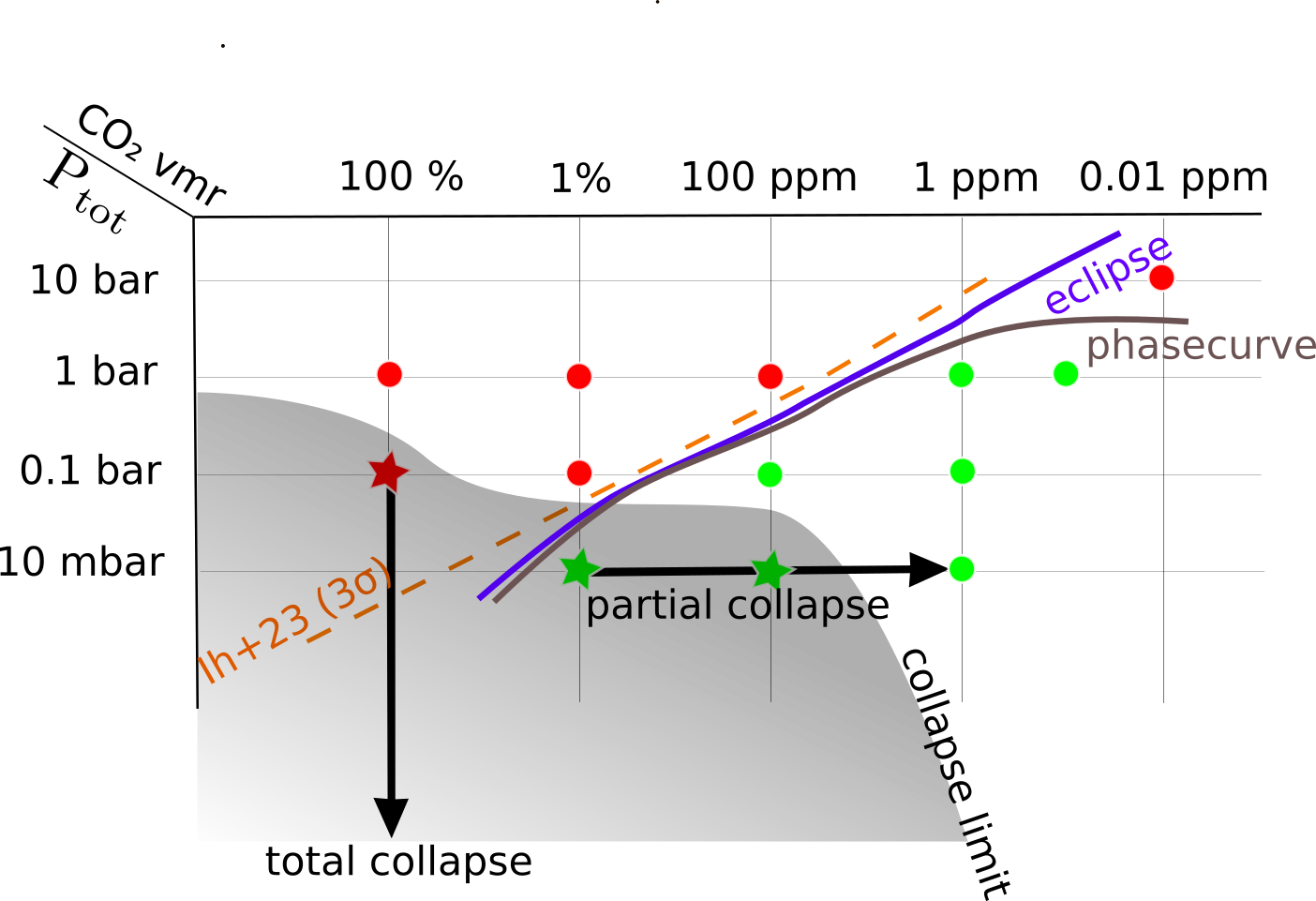}
    \caption{N$_2$-CO$_2$ grid of performed simulations. The points in red are not compatible with the observations; the green points match the data. The orange, purple, and brown lines correspond respectively to the 3$\sigma$ limit of \cite{Ih:2023}, and limits from this work from eclipse and phase curve observations. The arrows indicate a collapse of the CO$_2$ present in the atmosphere. The dots correspond to stable atmosphere, stars to unstable ones. The gray area indicates the atmospheres that are unstable and will collapse.}
    \label{fig:N2-CO2-grid}
\end{figure}

\subsubsection{Constraints from observations}
\begin{figure*}
    \centering
    \includegraphics[width=\linewidth]{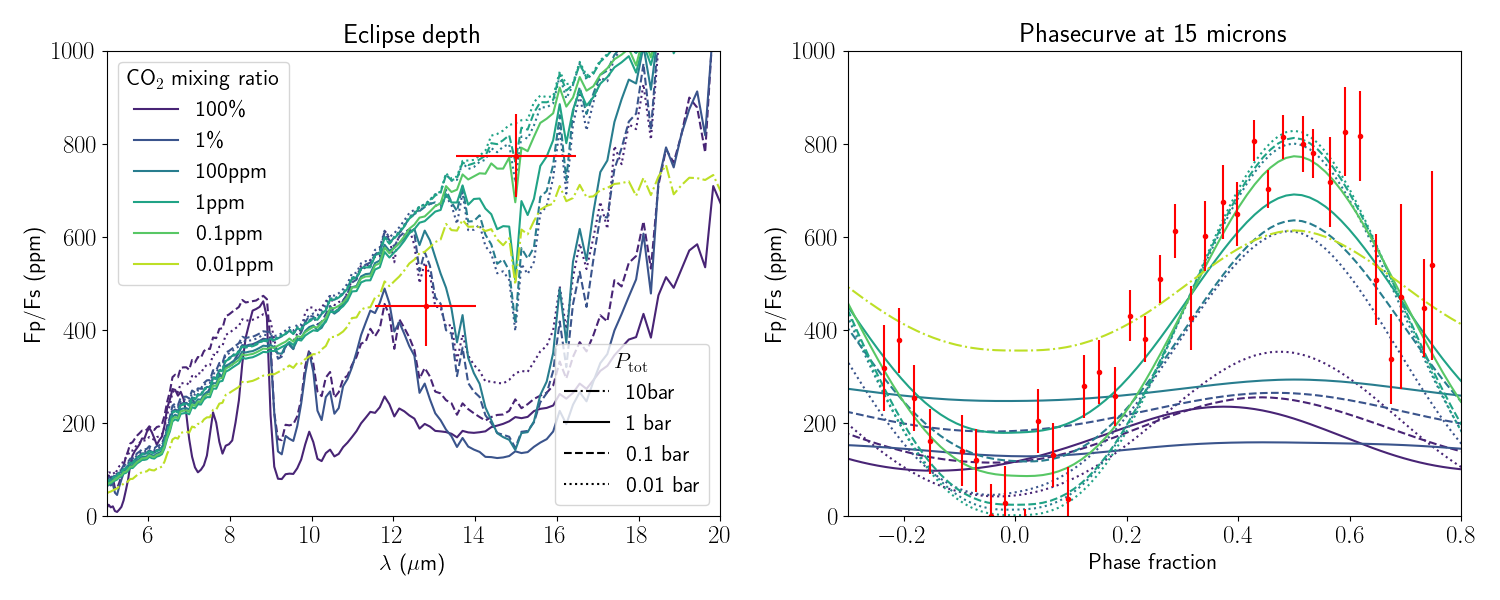}
    \caption{Synthetic eclipse spectra (left panel) and phase curves (right panel) for all the CO$_2$-N$_2$ atmospheres.}
    \label{fig:N2-CO2-obs}
\end{figure*}

Fig.~\ref{fig:N2-CO2-obs} shows the observables resulting from the simulations of the grid. In order to compare them to the observations, we computed the reduced $\chi^{2}$, provided in Table~\ref{table:method_simulations_list}.

\subsubsection{Eclipse depths}\label{sec:results/co2/obs/eclipse}
For the eclipse depths, the $\chi^2$ is high for all the high-P${_{\mathrm{CO}_2}}$, and drops for the atmospheres that correspond to the 3$\sigma$ limit of \cite{Ih:2023} (Fig.~\ref{fig:N2-CO2-grid}). It corresponds to all the cases that match the 15~$\mu$m data point within the 1$\sigma$ error bar in Fig.~\ref{fig:N2-CO2-obs}. We find that the eclipse depth is mostly dependent on the CO$_2$ partial pressure, and could not distinguish the different thicker ``transparent'' atmospheres with constant CO$_2$ partial pressure but variable total pressure.

In this study, the emission for an airless planet of albedo 0.2 is higher than in \cite{Ducrot:2024}. In their work, the albedo is assumed to be wavelength-independent, with the emissivity defined as $(\epsilon_\lambda = 1 - A)$. As a result, the surface temperature, governed by the equation $(1-A) ISR=\epsilon \sigma T^4$ (where ISR represents the incoming stellar radiation) becomes independent of the albedo, and the emission is directly proportional to the emissivity, (1-A). In contrast, with \texttt{exo\_k}, we impose an albedo cutoff at $\lambda \geq 5\mu$m, which results in an emissivity $\epsilon_\lambda=1-A_\lambda$ equal to 1 in thermal wavelengths. $\lambda=5 \mu$m was chosen to separate the zone of stellar emission and the zone of thermal planetary emission (Fig.~\ref{fig:1D-inversion}). This leads to a surface temperature that depends on the albedo. Although the bolometric planet emission remains proportional to (1-A), thus maintaining the radiative balance, the emission at 15~$\mu$m is not. Finally, for the 3D simulations, the albedo is assumed to be wavelength-independent, but the emissivity is set to 1, which gives similar results to \texttt{exo\_k}, as the albedo at longer wavelengths does not significantly influence the radiative balance, due to the low stellar emission at these wavelengths, and conversely, the emissivity is important only in the thermal wavelengths.

In consequence the surface albedo of 0.1, corresponding to the assumption made in simulations presented in \cite{Ducrot:2024}, results in a high eclipse depth for the thinnest atmospheres. The complexity and time-consumption of the GCM simulations does not allow us to explore widely the effect of the surface albedo, but we tested it in 1D for the different cases. (see Appendix~\ref{sec:app-albedo}). We found that a higher albedo lowers the continuum part (where we probe the surface), but not the CO$_2$ band when it is strong, because in this case we probe the emission directly from the atmosphere. We thus consider that the match with the data would be better with a higher albedo, but it would not change our overall conclusions.

\subsubsection{Phase curve}\label{sec:results/co2/obs/phasecurve}
From the phase curve in the 15~$\mu$m band, as expected, the atmospheres that present a strong CO$_2$ feature in eclipse give a very low dayside flux. As we are probing in the atmosphere because of the CO$_2$ absorption, the redistribution is visible in these atmospheres through the non-zero nightside flux, and even a strong offset for the 1~bar 100\% CO$_2$ case. These atmospheres are thus ruled out again, as expected.

The thinner atmospheres, with a higher dayside flux, low nightside flux and zero peak offset, can be considered an acceptable match, corresponding to the analysis in eclipse.

The only case where the phase curve brings a new constraint is for the thicker, more transparent atmosphere (see Appendix~\ref{sec:app-transparent_atmos}). In this case, even with a very low quantity of CO$_2$, the weak continuum absorption (CIA and continua) is enough for the atmosphere to absorb and re-emit in the nightside . The probed atmosphere is well enough mixed that the nightside flux is high, and thus, not compatible with the observations.

\subsubsection{Constraints from atmospheric collapse}\label{sec:results/co2/collapse}

TRAPPIST-1 b is so close to its host star that it is very likely to be tidally locked \citep{Turbet:2020review}. Therefore, in the absence of atmosphere, its dayside is expected to be very hot and its nightside very cold. An atmosphere can redistribute the heat from the dayside to the nightside, with an efficiency that depends on its thickness, composition and, in general, on the parameters that can impact large-scale dynamics \citep{Wordsworth:2015,Koll:2022}.

When reaching its condensation temperature, the gaseous CO$_2$ is expected to condense and form clouds, and/or frost on the ground. Therefore, if the nightside is cold enough, it will constitute a cold trap, collapsing the CO$_2$ of the atmosphere onto the surface \citep{Wordsworth:2015,Turbet:2018aa}.

The \texttt{Generic PCM} simulations we performed include CO$_2$ condensation, allowing us to test this prediction. We thus used the model to explore for which cases (pressure, CO$_2$ mixing ratio) the collapse happens, and if the collapse is total or if a secondary equilibrium is achieved \citep{Soto:2015}.

\begin{figure}
    \centering
    \includegraphics[width=\linewidth]{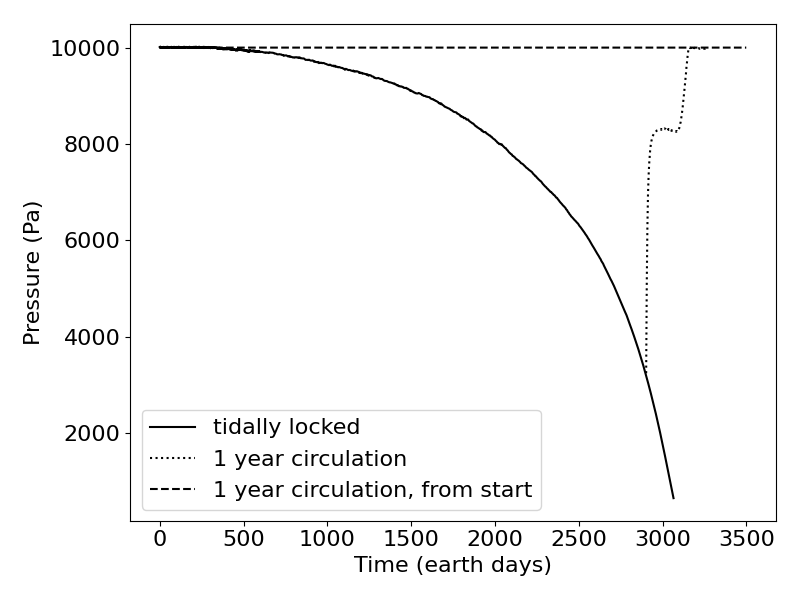}
    \caption{Total pressure of a pure CO$_2$ atmosphere against time, for the threshold case of a 0.1~bar atmosphere. In a tidally locked case, the atmosphere collapses totally in approximately ten Earth years. If we set a circulation of the substellar point to one Earth year, the 0.1 bar atmosphere becomes stable, and a partially collapsed atmosphere vaporizes all the CO$_2$ back to the atmosphere.}
    \label{fig:collapse_vs_time}
\end{figure}

\paragraph{{Details on a pure CO$_2$ simulation collapse}}
We find that for a pure CO$_2$ case, the atmosphere collapses as soon as the initial pressure is 0.1~bar or below. This result is not impacted by the number mixing ratio of cloud condensation nuclei and radiative effect of CO$_2$ ice clouds. For this end member case, we performed the simulation of the collapse down to a few pascals (Fig.~\ref{fig:collapse_vs_time}, solid black line), demonstrating that there is no secondary equilibrium possible. We computed the observables at different steps of the collapse, that are displayed in Appendix~\ref{sec:app-co2}.

Simulations showed that the ice is mainly produced as a result of frost formation and not precipitation. We note that it implies that the collapse strongly depends on the temperature of the surface at the cold traps, which has been shown to be strongly dependent on the strength of turbulence on the nightside  \citep{auclair:2022, joshi:2020}.

Due to global dynamics, the cold traps are located close to the terminator, at mid-latitudes.
As the atmosphere collapses, the greenhouse effect and redistribution become less efficient, the surface temperature drops, and the cold traps become more and more pronounced.

We tested the impact of the radiative effect of CO$_2$ ice aerosols parametrization. Changing the ratio of the cloud condensation nuclei and turning off the radiative effect of the aerosols did not change our results.

\paragraph{{Collapse for a N$_2$-CO$_2$ atmosphere}}

For the N$_2$ atmospheres with 1\% to 1 ppm of CO$_2$, partial pressure of CO$_2$ is lower so its condensation temperature decreases, and the N$_2$ participates in the redistribution of heat. Thus, the collapse is prevented for 0.1~bar atmospheres. For thinner atmosphere, when these processes are not sufficient to avoid collapse, CO$_2$ condenses, but not N$_2$. We called it a partial collapse (Fig.~\ref{fig:N2-CO2-grid}). In this case, the total pressure does not drop, but the CO$_2$ gas content drops. These atmospheres are thus collapsing towards an atmosphere with fewer CO$_2$ (for 0.1~bar total pressure, 1ppm is stable) eventually producing a transparent-type atmosphere, in better agreement with the JWST MIRI observations. Therefore, the case of a thin, N$_2$ atmosphere with a small amount of CO$_2$, which matches available data, can be considered as a resulting atmosphere of the CO$_2$-richer atmospheres, and not as implausible as it would look at first sight.

\subsubsection{Processes that can stop atmospheric collapse}

\paragraph{Non purely synchronous orbit}
Like a lot of exoplanets, TRAPPIST-1 b is close enough to its star to be expected to be in synchronous rotation due to tidal interaction with the star. But the TRAPPIST-1 system is so compact that planets gravitationally interact also with each other and this could result in a movement of the substellar point: libration, and a slow circulation of the substellar point. The circulation, if not too slow, can prevent collapse of the atmosphere by re-sublimating the ice caps when exposed to starlight. A 3:2 spin-orbit resonance could also prevent the collapse, but for TRAPPIST-1 b the tidal effects make it improbable \citep{Turbet:2020review,Revol:2024}.
We tested this hypothesis by simulating a 0.1~bar atmosphere of pure CO$_2$ (collapsing in the tidally locked case), with a substellar point circulation (i.e., synodic day) of 1 Earth year. A longer duration would have been interesting to investigate but too computationally expensive and time-consuming. First, we started like the tidally locked case, with a warm (500~K), isothermal, dry atmosphere, and in another case, we started from a partially collapsed atmosphere. In both tests, the reached equilibrium state was a dry, non-collapsed atmosphere.
The timescale for the collapse of a 0.1 bar atmosphere of pure CO$_2$ being around 10 Earth years (Fig.~\ref{fig:collapse_vs_time}), a circulation of the substellar point with a rotation period of 1 Earth year is enough to prevent collapse. We can expect that the characteristic time of the substellar drift must be not longer than the one of the collapse in order to prevent it.
The complete rotation of the substellar point has already been suggested by several studies.
\cite{Makarov:2018} suggested possible spin-orbit resonances state in the spin evolution of the planets b, d, and e, before synchronization.
Chaotic circulation of the substellar points has been explored by \cite{Vinson:2019}, \cite{Shakespeare:2023}, and \cite{Chen:2023}.
They attribute the large librations rising from the chaotic motion in their simulations, coupled with a simplistic model of tides (e.g., Constant Time Lag model \citep{Hut:1981}).
In addition, \cite{Revol:2024} computed the synodic periods for each of the TRAPPIST-1 planets, finding a synodic day of about 70 Earth years for planet b.

In the end, for TRAPPIST-1 b, escaping atmospheric collapse thanks to the circulation of the substellar point is quite unlikely since simulations show that its circulation is not expected to be faster than one complete rotation of the substellar point per $\sim$ 70 Earth years \citep{Revol:2024}. However, the circulation of the substellar point is expected to be more important for the outer planets and should be taken into account for future GCM studies.

\paragraph{{Tidal heat flux}}

On TRAPPIST-1 b, the star-planet tidal interactions may induce an internal flux. The eccentricity from \cite{Agol2021} constrains this flux between $4 \times 10^{-2}$ and 500 W.m$^{-2}$. An internal heat flux can impact CO$_2$ collapse in two ways: 1) if it is strong enough, it can prevent any condensation at the surface, and 2) it can limit the height of the ice caps because of the basal melting, thus limiting the amount of CO$_2$ that can be condensed at the surface.\\
First, we can compute the equilibrium temperature at the nightside induced by an internal heat flux:
$$\sigma T_\mathrm{eq}^4 = F_\mathrm{int}.$$
If this temperature is superior to the condensation temperature of CO$_2$, CO$_2$ condensation can not occur, and thus no collapse. Thus, we can determine the minimum partial CO$_2$ pressure, corresponding to the pressure at which the condensation temperature is equal to the computed equilibrium temperature.\\
Second, the CO$_2$ ice cap can be limited by basal melting \citep{Turbet:2017epsl}. The maximum volume of CO$_2$ ice is limited to an ice cap of the surface of the nightside and the maximum height fixed by basal melting. Thus, the maximum pressure of CO$_2$ that can condense corresponds to
$$P_{base}=\frac{1}{2}g\rho h_{max}$$
with $g$ the gravity of the planet, and $\rho$ the density of CO$_2$ ice.

We computed for different orders of magnitudes of the internal flux, the minimum pressure of CO$_2$ from 1), and the maximum condensed CO$_2$ from 2). Results are in Table~\ref{tab:int_flux}. 

From this table, we identified three regimes: \\
(a) For a flux of $\sim$ 100W.m$^{-2}$ and higher, the condensation pressure is higher than the boundary pressure at which the atmosphere starts collapsing $P_{collapse}$. Therefore, there is no collapse possible.\\
(b) For a flux of $\sim$ 1W.m$^{-2}$ and lower, the internal flux does not prevent condensation, and the maximum amount of CO$_2$ trapped correspond to a CO$_2$ pressure higher than $P_{collapse}$, and thus there is no constraint brought by the internal flux.\\
(c) For intermediate cases around 10W.m$^{-2}$, process (2) tells us that the amount of CO$_2$ condensed is limited to $\sim$ 0.1 bar, which is below $P_{collapse} \sim$ 0.3 bar (for a pure CO$_2$ atmosphere). Thus, if there is an initial pressure between 0.1 bar and 0.3 bar, the atmosphere will not collapse completely, and there could be a remaining atmosphere, thinner than the initial limit $P_{collapse}$.

\begin{table}
    \centering
    \caption{Tidal heat flux constraints.}
    \begin{tabular}{c|c|c|c|c}
       F$_{int}$  & T$_{eq}$  & P$_{cond}$ & h$_{max}$  & $P_{base}$ \\
       (W.m$^{-2}$) & (K) &  (bar) & (m) & (bar) \\
       \hline
        100 & 204 & $\sim$ 1 & -- & --\\
        10 & 115 & $\sim$  1.e-4 & $\sim$ 1 & $\sim$ 0.1 \\
        1 & 64 & -- & $\sim$ 10 & $\sim$  1\\
        0.1 & 36 & --  & $\sim$ 100 & $\sim$ 10\\
    \end{tabular}
    \tablefoot{For different internal heat flux  F$_{int}$, the equilibrium temperature of the nightside T$_{eq}$ and the corresponding equilibrium vapor pressure  P$_{cond}$, the maximum height of the ice cap h$_{max}$ and the maximum pressure of CO$_2$ that can condense before saturating the ice caps $P_{base}$.}
    \label{tab:int_flux}
\end{table}

\subsection{Thick greenhouse-efficient atmosphere}\label{sec:results/thick}

\subsubsection{Principle}\label{sec:results/thick/pple}

This family of atmospheres represents high-redistribution, high-greenhouse effect cases. To produce an efficient greenhouse effect, the gases present in the atmosphere must have efficient absorption in the thermal infrared, where the planet emission is strong. This way, the atmosphere should prevent the thermal radiation to directly escape to space. In other words, it means that the atmosphere becomes optically thick very high in the atmosphere, producing a brightness temperature representative of the higher, likely colder atmospheric layers, and not of the hot surface and/or deep atmosphere. The more absorbent the atmosphere is, the higher atmospheric layers we  probe in secondary eclipses, and the colder the brightness temperature we  see.

In principle, this family of atmosphere should therefore be incompatible with the measured secondary eclipses at 12.8 and 15~$\mu$m. However, this family of atmosphere can theoretically generate a very high flux at the wavelengths measured by MIRI, if these wavelengths coincide with atmospheric absorption windows. This also requires that the atmosphere does not absorb too efficiently in the near infrared to let the stellar flux reach the deep atmosphere. Fig.~\ref{fig:thick_1D_fke_op} shows a proof-of-concept example where we set up an atmosphere for TRAPPIST-1 b with artificial opacities (high opacity set in the thermal infrared, but low opacities near 12.8 and 15~$\mu$m). This thought experiment, detailed in Appendix~\ref{sec:app/thick/concept},  is designed here simply to illustrate the mechanism at play.

\subsubsection{Gases with the right spectroscopic properties}\label{sec:results/thick/comp}

We looked for a composition that could produce such an effect. We do not focus on the plausibility of the composition, as there are still few constraints regarding planetary evolution. We allow ourselves the freedom to explore all compositions, with the sole criterion being compatibility with the data. Among a list of possible molecules (CH$_4$, CO, CO$_2$, C$_2$H$_2$, C$_2$H$_4$, C$_2$H$_6$, HCN, H$_2$O, H$_2$, H$_2$S, H$_2$O$_2$, NH$_3$, N$_2$, PH$_3$, SO$_2$) we selected the species that had a weaker absorption around 15~$\mu$m. C$_2$H$_4$ and C$_2$H$_6$, i.e., reduced gases and potential photochemistry products of CH$_4$, are the only molecules with such a feature (Fig.~\ref{fig:molec_opacities}). Thus, to present this mechanism, an atmosphere must likely be reduced, and contain at least one of these gases. Other gases, transparent or with limited greenhouse effect, can be present in addition.

After an exploration in 1D with \texttt{exo\_k} with these two gases and N$_2$ and CH$_4$, we identified a composition that maximized the dayside emission: N$_2$ as the background gas, CH$_4$ (20\%) and C$_2$H$_4$ (0.2\%). This composition has been chosen for the 3D \texttt{Generic PCM} simulations. We tried this composition with a total pressure of 10~bar.

We note that a larger quantity of CH$_4$ would make the atmosphere absorb too much in the near infrared, where TRAPPIST-1 emits a significant fraction of its flux, and act as an anti-greenhouse effect, preventing the stellar flux to go deep in the atmosphere (see Section~\ref{sec:results/inversion}).

\subsubsection{Constraints from the secondary eclipse}\label{sec:results/thick/eclipse}
We show in Fig.~\ref{fig:obs_thick} the results and observables of the simulation of N$_2$-CH$_4$-C$_2$H$_4$, with P$_\mathrm{tot}$~=~10~bar. Due to the strong greenhouse effect, the surface of the planet is quite hot: more than 600~K (compared to the equilibrium temperature of $\sim$400~K). The thermal profiles of the dayside and nightside are superimposed in the deep atmosphere, because of the highly efficient redistribution. In the upper layers, the redistribution is less efficient, and the thermal profiles of dayside and nightside diverge. The layers from which the main part of the 15~$\mu$m band pass emission comes from (shown in shadowed area) are between 1~bar and 0.01~bar, because even with the opacity ``window'', the atmosphere becomes optically thick due to the opacity scaling with atmospheric pressure, and one cannot probe all the way down to the surface. Thus, the eclipse depth is around 600 ppm for 15~$\mu$m, corresponding to a brightness temperature of only $\sim$425~K, and 450ppm / $\sim$415~K for 12.8~$\mu$m.
\subsubsection{Constraints from the phase curve}\label{sec:results/thick/phasecurve}

\begin{figure*}
    \centering
    \includegraphics[width=\linewidth]{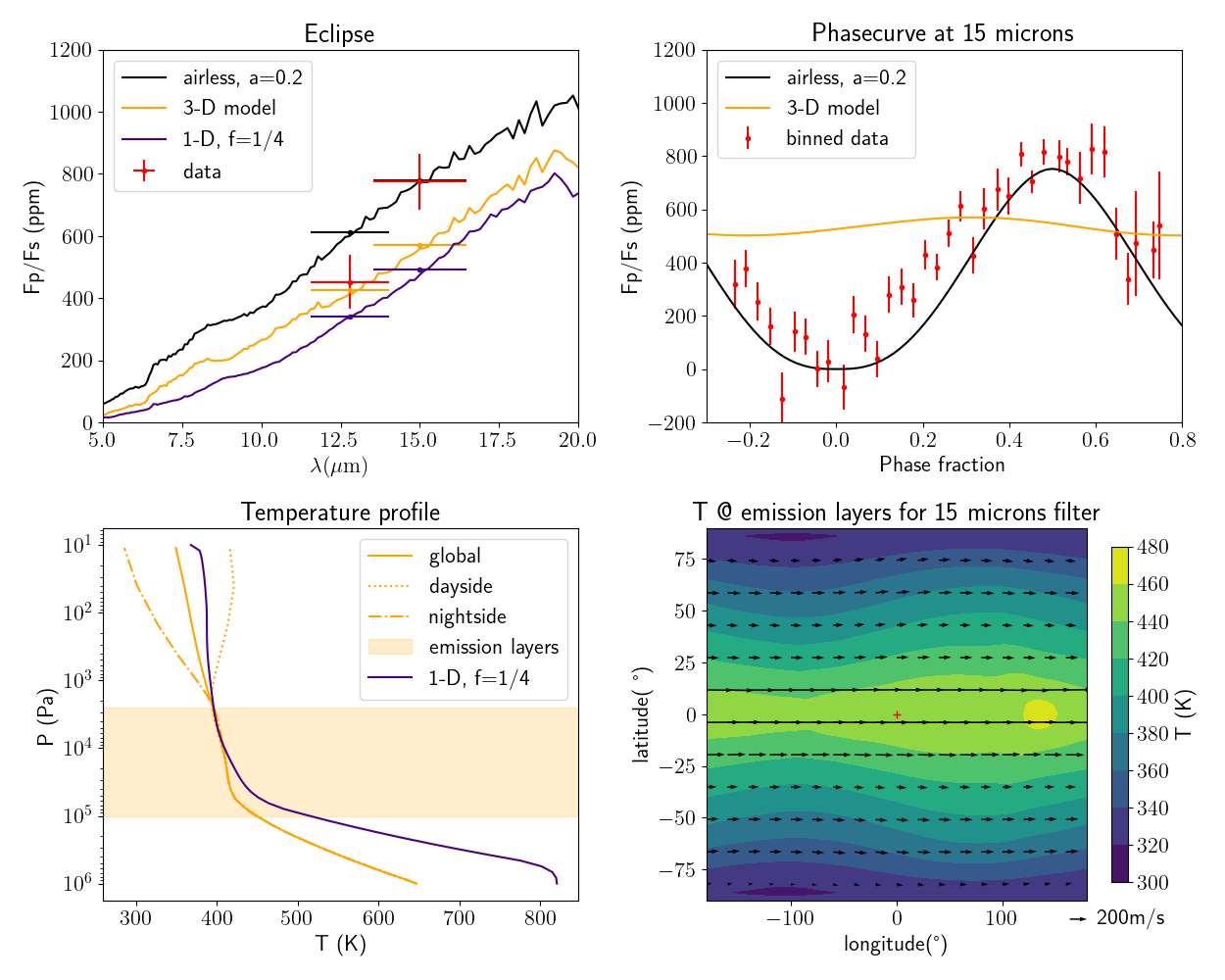}
    \caption{Thick atmosphere case (10~bar, N$_2$ atmosphere with 20\% of CH$_4$ and 0.2\% of C$_2$H$_4$). Upper panels: Synthetic observables and observed data (eclipse depth vs wavelength, left, and phase curve, right). Lower panels: Temperature profile (left) and map (right). The temperature map corresponds to a sum of the atmospheric layers, weighted by their contribution to the emission at 15~$\mu$m. We added the similarly weighted wind vectors as black arrows. The shadowed area in the right panel corresponds to the layers contributing to 95\% of the emission. The difference between the 1D and 3D temperature profiles is discussed in Appendix~\ref{sec:app/thick/3D}.}
    \label{fig:obs_thick}
\end{figure*}

Still in Fig.~\ref{fig:obs_thick}, the wind and temperature map at the emission layers shows strong eastwards winds, moving heat towards the nightside of the planet. These winds have also the effect of moving the hottest point eastwards the substellar point.

The modeled phase curve reproduces these features: it is globally flat, because of heat redistribution, and the maximum of the phase curve is shifted slightly ahead of the time of the eclipse, because of the shift of the hottest point. In conclusion, such an atmosphere is compatible with the eclipse depth measurements, but is ruled out by the phase curve.

\subsection{Atmospheres with a strong thermal inversion}\label{sec:results/inversion}
\subsubsection{Principle}\label{sec:results/inversion/pple}

In first approximation, Earth's atmosphere is transparent in the visible, and opaque in the IR, allowing most of the light from the Sun to go through all the atmosphere, warming the surface, which re-emits this energy in the infrared, and the atmosphere itself absorbs this light. This process tends to make the atmosphere warmer close to the surface, and colder as we go up in the atmosphere.
On Earth, the ozone layer absorbs in the UV, heating the atmosphere directly from the top. It creates a so-called thermal inversion: the atmosphere goes hotter as we go up, in what is called the stratosphere.

In the case of TRAPPIST-1 b, the star is colder than the sun, and its spectrum is thus shifted towards infrared wavelengths. A significant part of the emission of the star is in the near infrared. Thus, a strong thermal inversion can happen quite easily, if the atmosphere contains some kind of absorber in the stellar emission range. Aerosols, but also molecules like CH$_4$, can absorb in the near infrared. In this section, we explore thermal inversion created by different kinds of components.

\begin{figure*}
    \centering
    \includegraphics[width=\linewidth]{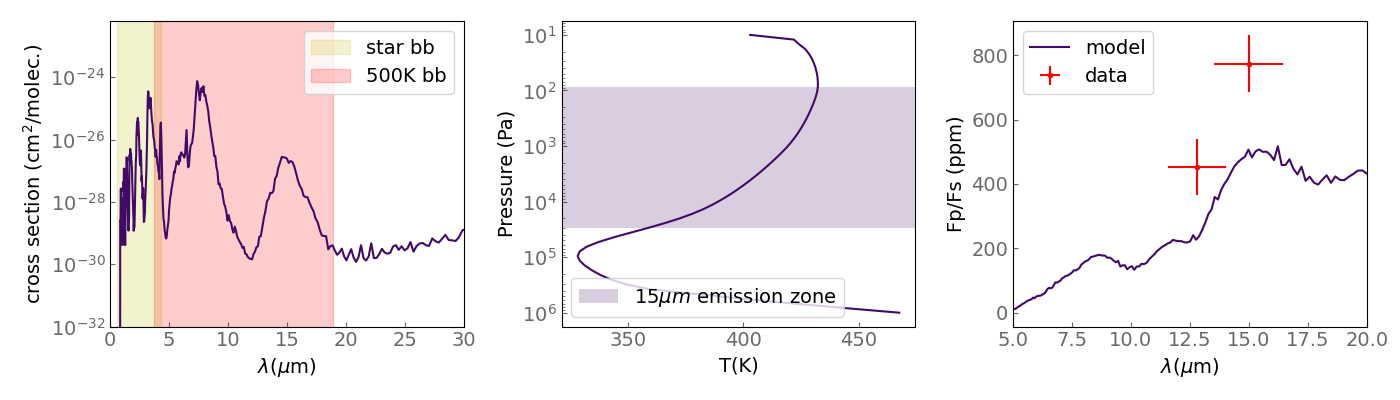}
    \caption{1D case of thermal inversion (10 bar, N$_2$+40\% CH$_4$+0.4\%CO$_2$). Left: Opacity of the atmosphere with the zone of emission of the star and the planet. Middle: Temperature profile and the layers of emission of the atmosphere at 15~$\mu$m. Right: Eclipse depth. The methane absorbs partly in the star-emission region, creating the thermal inversion in the temperature profile. The CO$_2$ absorbs in the 15~$\mu$m band, and thus the emission layers at 15~$\mu$m are high enough in the atmosphere, where the temperature is high. Thus, the eclipse depth is higher.}
    \label{fig:1D-inversion}
\end{figure*}

If there is a thermal inversion, a molecule with an absorption band at the observed wavelength will result in observations probing higher in the atmosphere, as usual, but this time, it means that one will probe higher temperatures. The absorption band of the molecule is thus expected to be observed as an emission band (Fig.~\ref{fig:1D-inversion}).

\begin{figure*}
    \centering
    \includegraphics[width=\linewidth]{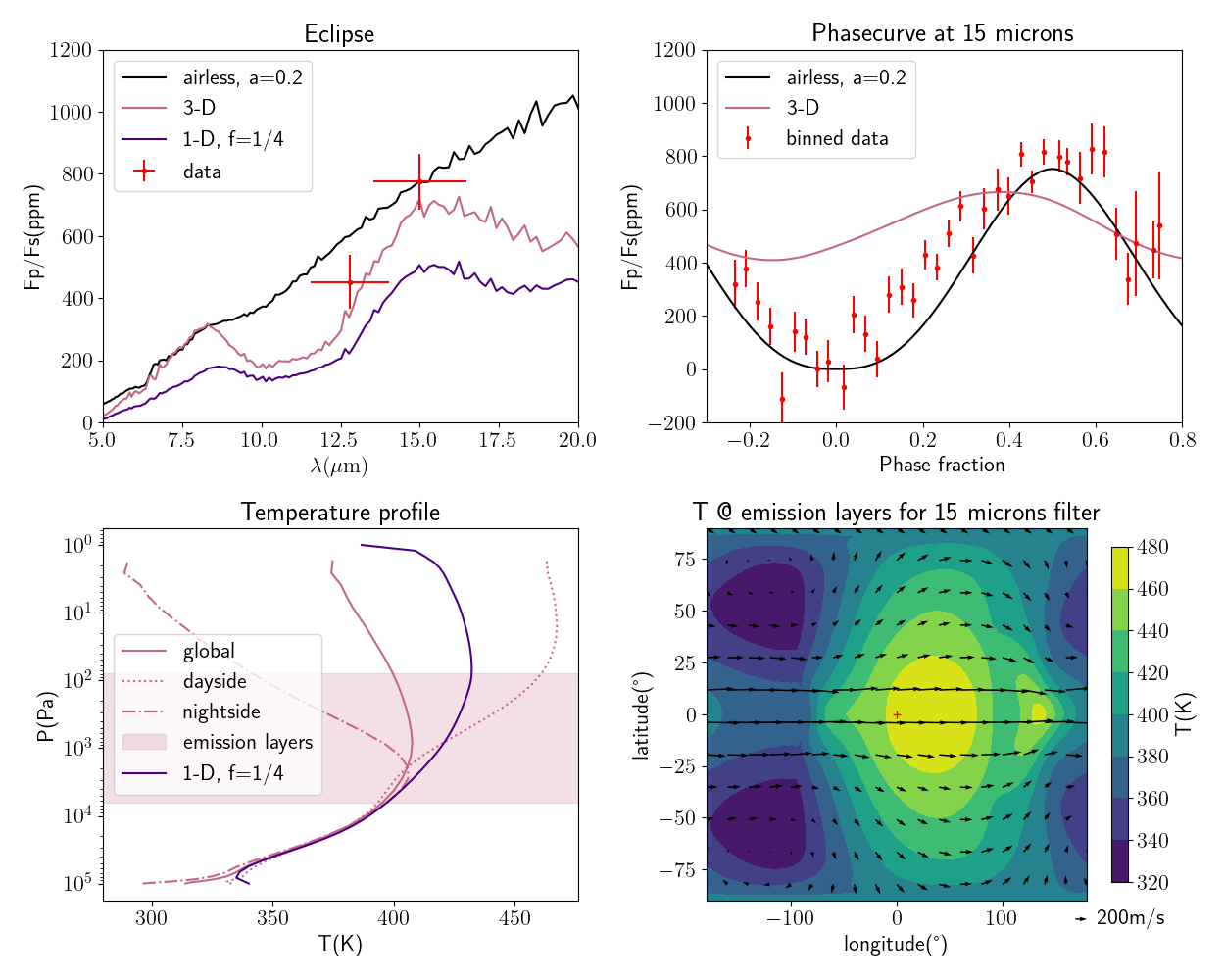}
    \caption{Thermal inversion: 1 bar, N$_2$+40\% CH$_4$+0.4\%CO$_2$. The panels are the same as in Fig.~\ref{fig:obs_thick}. }
    \label{fig:obs_CH4}
\end{figure*}

\subsubsection{Composition}\label{sec:results/inversion/comp}
\begin{figure*}
    \centering
    \includegraphics[width=\linewidth]{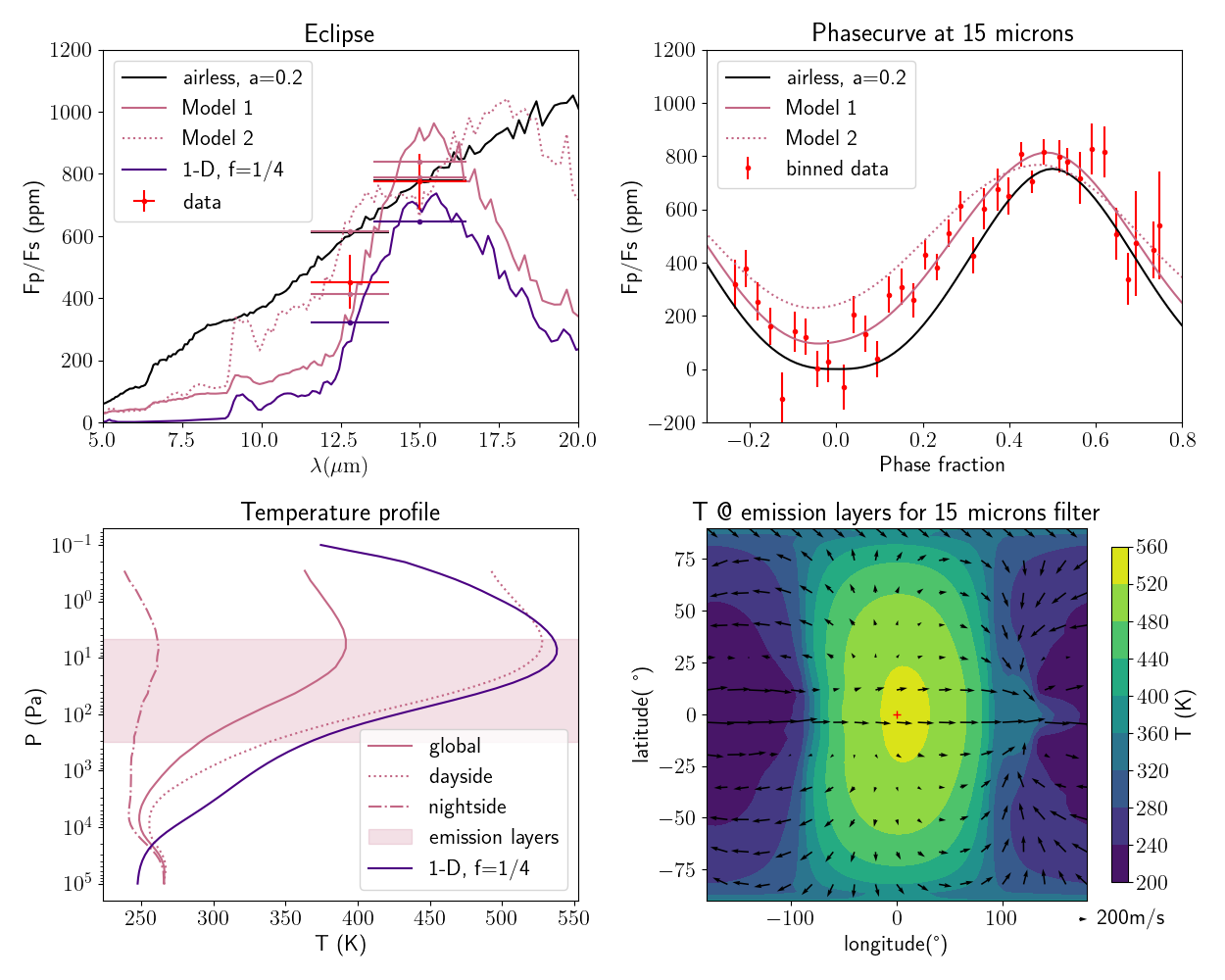}
    \caption{Thermal inversion: CO$_2$+ simplified hazes. The panels are the same as in Fig.~\ref{fig:obs_thick}. Model 1 and Model 2 for the hazes are described in the text. The temperatures and winds of the lower panels correspond to the Model 1 case, which is the model closest to the data.}
    \label{fig:obs_haze}
\end{figure*}

\paragraph{{Thermal inversion with gas}}
We tested an atmosphere with a high amount of CH$_4$, to create the inversion, and some CO$_2$, to be seen in emission. We supposed a redistribution factor of 1/4 (total redistribution) for the 1D simulations. The results of the best case in 1D are shown in Fig.~\ref{fig:1D-inversion}. The chosen composition is N$_2$ as background gas, 40\% of CH$_4$, and 0.4\% of CO$_2$. The 3D study of this atmosphere is shown in Fig.~\ref{fig:obs_CH4}. The eclipse depth in 3D is much higher than in 1D, because the redistribution in the probed region is not total, contrary to the assumption made in 1D. We note that such an amount of CH$_4$ is expected to produce hazes in the upper layers, and that these hazes are expected to contribute to the thermal inversion.
\paragraph{{Thermal inversion with simplified hazes}}
Our simplified model for hazes, from \cite{Ducrot:2024}, is described in \ref{sec:methods/gPCM/hazes}.
We find that using the value of $f_\mathrm{haze}=7.0 \times 10^{-4}$ (Equation \ref{eq:haze_simple}) as in \cite{Ducrot:2024}, we probe a very high-altitude region of the atmosphere, where the heat redistribution is inefficient. Despite the strong winds, at such low pressure the radiative timescale ($\tau_\mathrm{rad}=\frac{C_pP}{g\sigma T^3}$ $\sim 10^2$s at P=10~Pa, with $C_p$ the heat capacity of the atmospheric gas, $g$ the gravity of the planet and $\sigma$ the Stefan-Boltzmann constant) is very short compared to the advection timescale ($\tau_\mathrm{adv}=\frac{R_p}{v} \sim 10^4$s, with $R_p$ the radius of the planet and $v$ the wind horizontal speed). Therefore, the dayside at this height is much hotter than the 1D prediction that assumed a full redistribution. In order to match the data we thus tried two different options: 1. Add a simplified single-scattering albedo of 0.5 by lowering the incoming stellar flux of a factor 2 (Model 1 , corresponding to \textit{Haze high} in \cite{ducrotgillon:2025}) and 2. Lower the haze factor $f_\mathrm{haze}$ to $f_\mathrm{haze}=3.0 \times 10^{-5}$, while adding a small simplified single-scattering albedo of 0.2 (Model 2, corresponding to \textit{Haze low} in \cite{ducrotgillon:2025}). These cases are shown in Fig.~\ref{fig:obs_haze}.

\paragraph{{Thermal inversion with more realistic aerosols}}
We also tested for aerosols for which we had more realistic optical properties: Titan-like tholins, H$_2$SO$_4$, and martian dust. A 1D study of the impact of the different parameters can be found in Appendix~\ref{sec:app/hazes/parameters}. For all aerosols tested, the single-scattering albedo was high in the near infrared, preventing a strong heating of the high atmosphere. Thus, a strong quantity of aerosols was needed to get a strong flux at 12.8 and 15~$\mu$m. The most efficient aerosols for the thermal inversion are the largest, and only the dust particles allowed for a strong flux at 15~$\mu$m compatible with the eclipse depth. For all the cases tested, the aerosols are very high and in large quantities, making the high atmosphere opaque at all wavelengths. At these low pressures, the CO$_2$ does not have a strong radiative effect, and is not seen in emission. Instead, the emission from the atmosphere corresponds to a blackbody at the temperature of the high atmosphere. We selected two cases to test in 3D: first, a layer of dust (with an effective radius of 100~$\mu$m) between pressures 1000 and 1~Pa, with a constant volume mixing ratio (VMR) of $3.5\times 10^{-18}$, and second, a layer of tholins (effective radius 1~$\mu$m, between 10 and 1~Pa, VMR=$1.\times 10{-13}$). We note that for the dust case, the opacity is so important that there is not enough flux to go through the atmosphere, resulting in an isotherm rather than a thermal inversion (see Appendix~\ref{sec:app/hazes/parameters}).

\subsubsection{Constraints from the secondary eclipse}\label{sec:results/inversion/eclipse}

For the inversion with gas and with simplified hazes, one can see that contrary to all other simulations, the CO$_2$ band is not seen in absorption, but in emission. This is consistent with the eclipse observations, since the 15~$\mu$m brightness temperature ($478\pm27$~K) is higher than the 12.8 one ($424\pm28$~K). For the inversion with realistic aerosols, the CO$_2$ is not visible as the atmosphere becomes optically thick too high.

For the simplified hazes case, the haze factor and single-scattering albedo were fit to obtain a satisfying match with the observations (Fig.~\ref{fig:obs_haze}). This case, although fine-tuned, gives the best agreement to the data (Table~\ref{table:method_simulations_list}).

\begin{figure*}
    \centering
    \includegraphics[width=\linewidth]{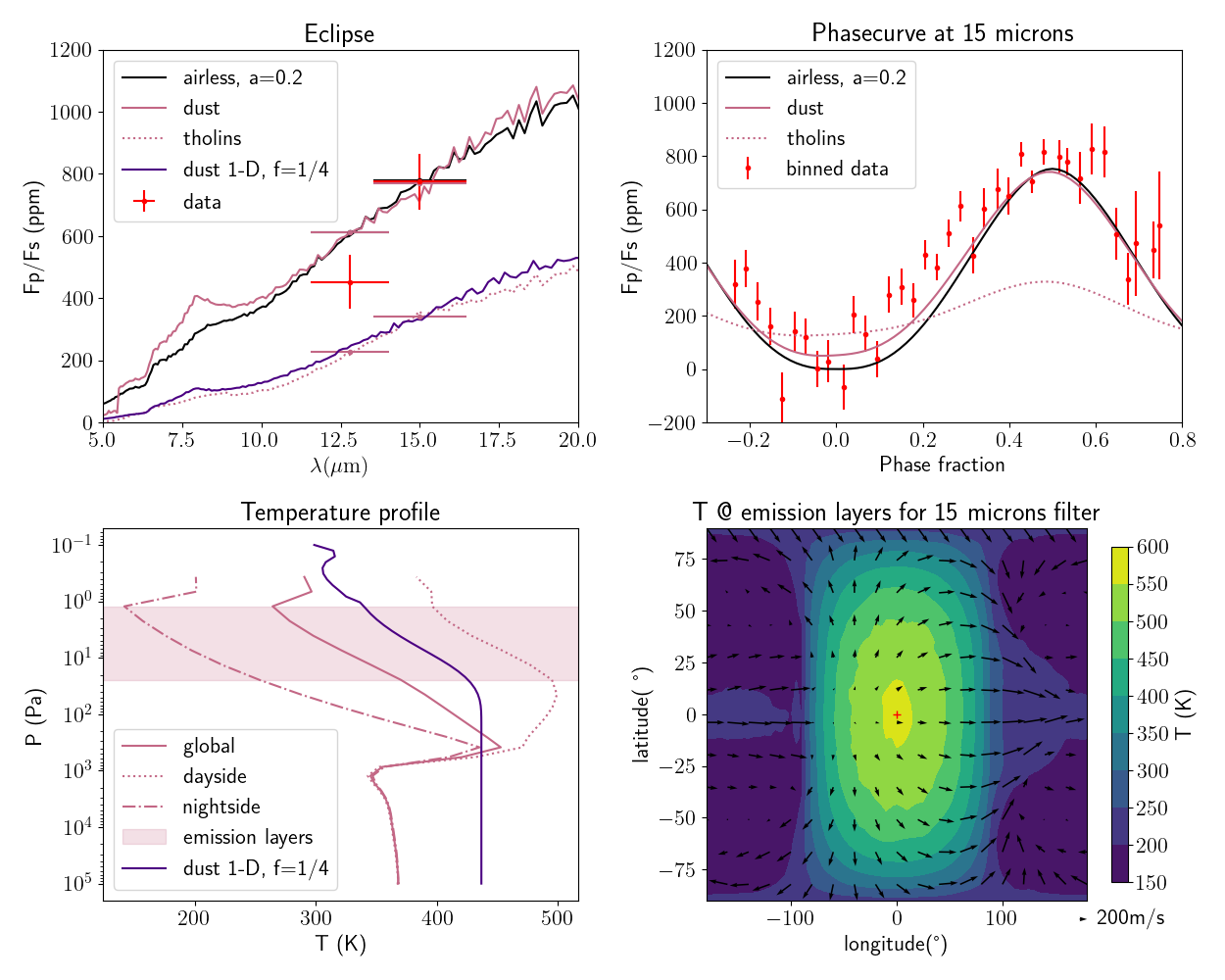}
    \caption{Thermal inversion: CO$_2$+ aerosols with realistic optical properties. The panels are the same as in Fig.~\ref{fig:obs_thick}. The parameters for dust and tholins are described in the text. The temperatures and winds of the lower panels correspond to the dust case, which is the model closest to the data. The difference between the 1D and 3D thermal profiles is discussed in Appendix~\ref{sec:app/hazes/convergence}.}
    \label{fig:obs_aerosols}
\end{figure*}

For both cases from aerosol optical properties (dust and tholins), the aerosol layer in the high atmosphere is so opaque that it mimics a surface, and the emission spectrum is similar to a bare rock (Fig.~\ref{fig:obs_aerosols}).

\subsubsection{Constraints from the phase curve}\label{sec:results/inversion/phasecurve}

For all cases, we probe high enough in the atmosphere to have a non-flat phase curve: despite the efficient mixing in the deeper atmosphere, the upper atmosphere presents a hotter dayside because the heat redistribution is less efficient there. 
For the CH$_4$ case (Fig.~\ref{fig:obs_CH4}) and the simplified haze Model 2 (Fig.~\ref{fig:obs_haze}), though, the probed pressures includes part of the well-mixed atmosphere, resulting in a low amplitude, high minimum temperature in the phase curve, but also a strong offset, because of the eastwards shift of the hottest point due to the super-rotation.

For the simplified haze Model 1 (Fig.~\ref{fig:obs_haze}) and more realistic aerosols, the probed region is so high, that all the redistribution is inefficient there, and the phase curve presents a quasi-zero nightside flux, a very high dayside flux, and no peak offset, thus matching the observations if the upper atmosphere is hot enough. Thus, to be compatible with the observations, the thermal inversion must be strong, and one needs a strong enough absorption at 15~$\mu$m to probe very high in the atmosphere. The hazes compatible are therefore abundant very high in the atmosphere and very dense. For the realistic aerosol optical properties (Fig.~\ref{fig:obs_aerosols}), the probed temperature is slightly too low to match the data. The realistic aerosols have a high single-scattering albedo, reflecting part of the stellar flux in space, so the upper layers temperature is lower than in the simplified cases, and the emission is weaker.

In summary, the emission signature is influenced by the type, abundance, and location of the absorbing species within the atmosphere. Three cases stand out: (1) When the stellar absorber is weak, such as CH$_4$, the emission originates from deeper atmospheric layers, with CO$_2$ detected in emission but resulting in a phase curve with low amplitude. (2) For more efficient absorbers, such as those in the simplified haze Model 1, the emission comes primarily from the upper atmosphere, leading to a phase curve with high amplitude mimicking the bare rock signal. (3) Finally, if the stellar absorber is highly efficient and abundant, it can render the entire atmosphere optically thick. Under these conditions, CO$_2$ will no longer be observed in emission, and the resulting emission signature will resemble that of an airless rock.

\section{Discussions}\label{sec:discussion}

\subsection{Limitations in the model}

The GCM allows for more realistic simulations, but they are too complex and expensive to pretend to explore the whole parameter space of exoplanet atmospheres. Among the parameters that we arbitrarily fixed and the hypotheses made, we note some that could have an impact on our results.
\begin{itemize}
    \item The internal heat flux of the planet, set to zero in our simulations, could prevent the atmospheric collapse for thin atmospheres, as already discussed in Section~\ref{sec:results/co2/collapse}
    \item The surface albedo changes the flux at 15 and 12.8~$\mu$m, in particular for optically thin atmospheres. We discuss its impact on the eclipse depth in Appendix~\ref{sec:app-albedo}.
    \item The hazes we added in our simulations are highly idealized: 
    we prescribed the vertical position and quantity of the hazes in the atmosphere. A more in-depth study of the formation of hazes would be necessary to assess whether the assumptions we made about haze properties are possible.
    \item Moreover, a CO$_2$-CH$_4$-rich atmosphere is not expected to be thermochemically stable \citep{Krissansen:2024,Watanabe:2024}. However, for this study we only tried to identify gas mixtures matching the data, without any prior on the expected composition. We recall that this is the first time as many observations are obtained on this category of planet (relatively temperate, Earth-size exoplanet around M-star), and leave the question of plausibility for future work.
    \item Apart from the N$_2$-CO$_2$ case, we did not study the formation and impact of clouds on our simulated atmospheres. At the pressure and temperature range studied, H$_2$O is the only condensable species. Condensation of water around TRAPPIST-1 planets have been studied already in \cite{Turbet:2023aa}.
\end{itemize}

\subsection{Limitations in the data analysis approach}

For a transit spectrum or a secondary eclipse, one fits the time-dependent data with a well-known light curve shape, and extracts the transit or eclipse depth from it.
The extraction of the phase curve is not as straightforward. The shape of the phase curve depends on the properties of the planet and its hypothetical atmosphere. Thus, one must make physical assumptions regarding the planet to parameterize the shape of the phase curve, so the data reduction is not as well decoupled from the physical interpretation as in the case of transits and eclipses.

Moreover, in our case,  the observed phase curve from \cite{ducrotgillon:2025} contains the contribution of TRAPPIST-1 b and c. Thus , it is delicate to extract the contribution of TRAPPIST-1 b alone in our study. The most relevant approach would be to perform a joint fit of the two planets, but this is out of the scope of this study.

\subsection{Comparison with previous studies}\label{sec:discussion/previous_studies}
\paragraph{{Lim et al, 2023}}
We computed synthetic transit spectra for the atmospheres that match all the data in emission, in order to check that these are compatible with the information available from transmission (see Appendix~\ref{sec:app/transit}). Our signal is typically below 100 ppm which is undetectable from the data of \cite{Lim:2023}. Thus, these scenarios can be neither ruled out nor verified from transmission spectroscopy information.\\
However, it is important to keep this in mind for future observations of TRAPPIST-1 system \citep{dewit:2024}, in particular observation strategies using TRAPPIST-1 b considered airless as a probe to correct for stellar contamination (JWST GO 6456).

\paragraph{{Ih et al, 2023}}
First, we compared our eclipse spectra from Fig.~\ref{fig:N2-CO2-obs} with the ones from \cite{Ih:2023} at same pressure and composition and obtained good agreement for atmospheres with little CO$_2$. For the atmospheres that are optically thick (e.g., pure CO$_2$ atmospheres), the flux in the CO$_2$ band is lower in our simulations, probably because of differences between our temperature profiles (the higher atmosphere is colder in our simulations).

Though we studied in depth the condensation of CO$_2$, we did not perform simulations for the other compositions in \cite{Ih:2023}. N$_2$, CO, O$_2$ and CH$_4$ should not condense under TRAPPIST-1 b conditions. SO$_2$ and H$_2$O, though, are more condensable than CO$_2$ (\cite{Fray:2009},\cite{Turbet:2018}), and \cite{Ih:2023} found that a 1~bar O$_2$ atmosphere with 100ppm H$_2$O and a 0.1~bar O$_2$ atmosphere with 100ppm SO$_2$ would fit the observations.  Considering the surface temperature in our 100ppm CO$_2$ simulations at these pressures, we find that the minimum surface temperature are below the condensation point of SO$_2$ and H$_2$O, and therefore most of the SO$_2$ and/or H$_2$O is expected to collapse for a tidally locked case.

\paragraph{{Ducrot et al, 2024}}
In \cite{Ducrot:2024}, the authors introduced the hypothesis of a thick, hazy atmosphere that allowed seeing CO$_2$ in emission. They expected an efficient redistribution for such a thick atmosphere. We find that with such an atmosphere, the redistribution is indeed efficient in the deep atmosphere, but in the higher, less dense atmosphere, which is actually the part that is probed at 15~$\mu$m, the heat redistribution becomes less efficient because of the very short radiative timescale at low pressure, allowing for a much brighter dayside, and a high amplitude phase curve (Fig.~\ref{fig:obs_haze}), an effect which was not represented by the 1D study.

\subsection{Perspectives for the study of TRAPPIST-1 b}

We highlighted different atmospheres that could produce observables compatible with our data available. Are there additional emission observations of TRAPPIST-1 b that could distinguish between the remaining cases ? The thin, residual atmospheres are by construction very similar to a bare rock, and cannot be distinguished by observation. For the atmospheres with thermal inversion, depending on the aerosols, we saw that  there can be a CO$_2$ emission band or no spectral features in emission. Only the case of a thermal inversion with CO$_2$ in emission could be investigated by more observations.

\begin{figure}
    \centering
    \includegraphics[width=\linewidth]{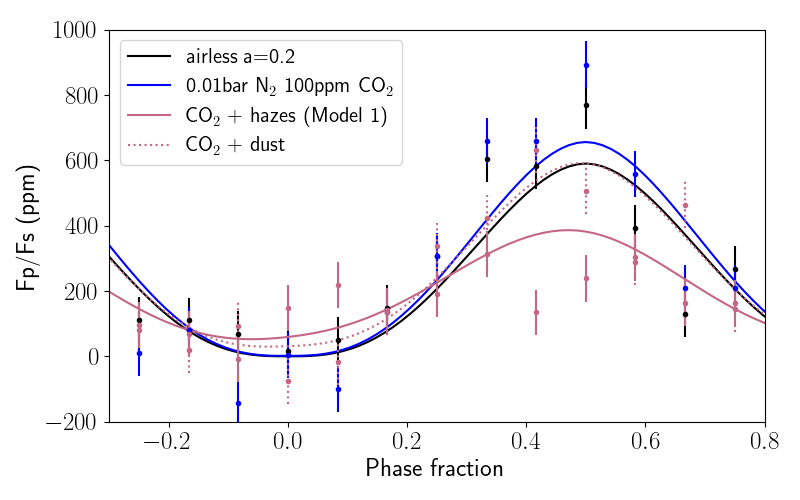}
    \caption{Phase curves in the 12.8~$\mu$m MIRI filter for an airless body and the remaining atmospheric families, along with the corresponding simulated observational points.}
    \label{fig:future_obs_phc}
\end{figure}

A phase curve at 12.8~$\mu$m could in principle help distinguish between some thermal inversion scenarios (with a lighter amplitude at 12.8~$\mu$m than 15~$\mu$m) and the airless case. However, the hazy atmosphere is still not transparent at these wavelengths, thus not resulting in a totally flat phase curve, and simulations of observations show that the signal-to-noise ratio would be too weak to distinguish the different scenarios (Fig.~\ref{fig:future_obs_phc}).

Secondary eclipse observations far from the CO$_2$ band, where some hazy models are expected to give much less flux than airless models, could be interesting. We used the JWST Exposure Time Calculator (ETC) to compute the expected precision on the occultation depth of TRAPPIST-1 b at 10 and 18~$\mu$m. Using the following formula we can compute the precision on the occultation depth:

\begin{equation}
    \mathrm{precision} = \frac{\sqrt{(1/N_{in} + 1/N_{out})}}{S/N_{point}}.
\end{equation}
Here $N_{in}$ is the number of points during the occultation and $N_{out}$ is the number of points outside the occultation, during a single visit.

\begin{figure}
    \centering
    \includegraphics[width=\linewidth]{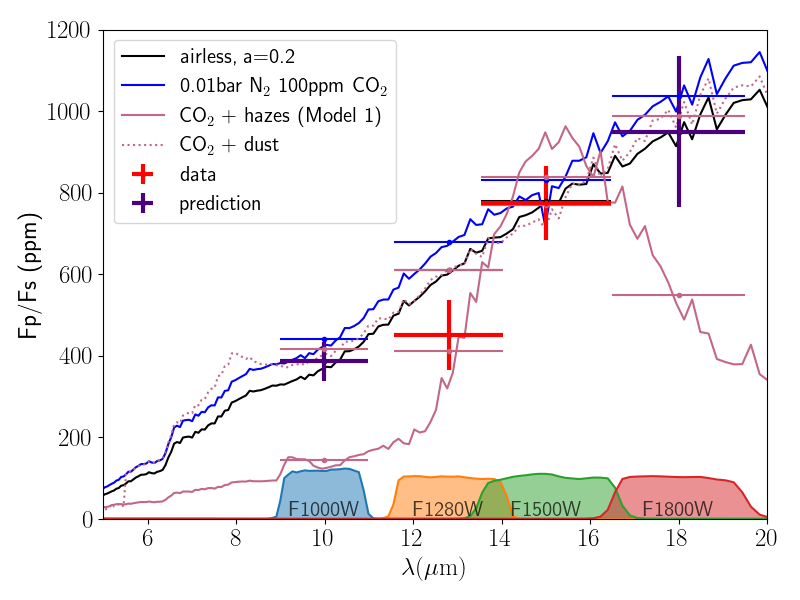}
    \caption{Dayside emission for an airless body and the remaining atmospheric families. The red points are from \cite{Ducrot:2024}. The purple points are predicted observation points for five visits for an airless body of albedo 0.2.}
    \label{fig:future_obs}
\end{figure}

For MIRI filter F1800W with SUB256 (to maximize the number of groups per integration) we obtain a S/N per point equal to 600 and an exposure time equal to 89.56s for 299 groups per integration.
We compute a precision of 371 ppm for one visit. For F1000W we compute a precision of 97ppm in one visit (S/N of 868.49 and exposure time of 13.18s). Fig.~\ref{fig:future_obs} shows the predicted observations for five visits. We compute that we would need at least eight visits to distinguish between a bare-rock scenario and the hazy, CO$_2$ emission case with an observation at 18~$\mu$m, but only two visits at 10~$\mu$m. This is to distinguish between two extreme cases, however, a hazy atmosphere emission can stand between these two cases, and even mimic an airless scenario (see Section \ref{sec:results/inversion/eclipse}).

\section{Conclusion}\label{sec:conclusion}

Starting from the secondary eclipse observations, we performed a study of the atmospheres on TRAPPIST-1 b that would produce a high brightness temperature at 12.8 and 15~$\mu$m. After exploring the parameter space in 1D, we found several families of atmospheres that do so, and selected representatives of each family to study in 3D. The 3D study brought several results that could not be predicted in 1D:

\begin{itemize}
    \item Thin condensable atmospheres: We identified the process of atmospheric collapse, through cold trap condensation, that allowed us to rule out some atmospheres that are thus not stable. Cross-checking this information with the observations allowed us to narrow the window of possible thin N$_2$-CO$_2$ atmospheres (Fig.~\ref{fig:N2-CO2-grid}). This result can be extended to other compositions (Section \ref{sec:discussion/previous_studies}).
    \item Thick transparent atmospheres: We found that even a very low opacity, such as N$_2$-N$_2$ collision-induced absorption, is enough to make the atmosphere radiatively active and emit a flux in the nightside, making such an atmosphere incompatible with the phase curve data (Appendix~\ref{sec:app-transparent_atmos}).
    \item Thick greenhouse-efficient atmospheres: We highlighted a type of thick highly redistributive atmosphere that gives a  dayside emission that is similar to bare rock, but that could be ruled out with the phase curve observation (Fig.~\ref{fig:obs_thick}). Atmospheres with these characteristics are reduced atmospheres (containing CH$_4$ and C$_2$H$_4$ or C$_2$H$_6$) in order to provide an atmospheric window for radiation to escape from deeper layers to give the high flux observed by JWST photometry.
    \item Atmosphere with thermal inversion: For this kind of atmosphere already proposed in \cite{Ducrot:2024}, we found that the 1D hypothesis of a full redistribution is not accurate since the flux is coming from high in the atmosphere where the heat redistribution becomes inefficient. Furthermore, this effect -- if the hazes have the right altitude and properties -- can result in a phase curve with a very strong day--night amplitude and no offset, compatible with the observations (Fig.~\ref{fig:obs_haze}). Only atmospheres with a thermal inversion and hazes or aerosols can match the phase curve data.
\end{itemize}

In summary, we showed that an airless planet and a thin residual atmosphere are not the only scenarios compatible with the secondary eclipse observations. We proposed three families of thick atmospheres (transparent atmospheres with almost no condensing or absorbing species; reduced atmospheres with an opacity window at 15~$\mu$m; and atmospheres with a thermal inversion) that also match the eclipse data. The phase curve observation showing a very cold nightside and no peak offset was needed to rule out some cases (thick reduced atmospheres; thick transparent atmospheres; and atmospheres with a thermal inversion but no hazes). Taking into account all the emission observations available, we showed that TRAPPIST-1 b might  be a bare rock or possesses a very thin atmosphere poor in absorbing and/or condensing gases; alternatively, it could possess an atmosphere with a thermal inversion and hazes, although this  case seems unlikely and fine-tuned.

Though this study was dedicated to TRAPPIST-1 b, as this planet had an unprecedented amount of observed data in emission (eclipse depths at two different wavelengths and phase curves), the methodology of this study is applicable to other rocky targets to be observed with JWST. Moreover, our results (e.g., atmospheric collapse, question of heat redistribution) can qualitatively apply to other close-in rocky exoplanets observed in emission. In particular, we call attention to the degeneracy of a high eclipse depth at 15~$\mu$m, making it hazardous to conclude that a planet is airless just from this photometric point.

\begin{acknowledgements} 
This work has made use of the Infinity Cluster hosted by Institut d’Astrophysique de Paris; we thank Stéphane Rouberol for his work on the cluster. We acknowledge support from the Tremplin 2022 program of the Faculty of Science and Engineering of Sorbonne University. We acknowledge support from the High-Performance Computing (HPC) resources of Centre Informatique National de l'Enseignement Supérieur (CINES) under the allocations No. A0140110391 and A0160110391 made by Grand Équipement National de Calcul Intensif (GENCI). We thank the Generic PCM team for the teamwork development and improvement of the model. We acknowledge support from BELSPO BRAIN (B2/212/PI/PORTAL).
\end{acknowledgements} 

\bibliographystyle{aa}
\bibliography{references}

\begin{appendix}\label{sec:appendix}
\onecolumn

\section{CO$_2$-N$_2$}
\subsection{Thick transparent atmosphere}\label{sec:app-transparent_atmos}
\begin{figure*}[h!]
    \centering
    \includegraphics[width=\linewidth]{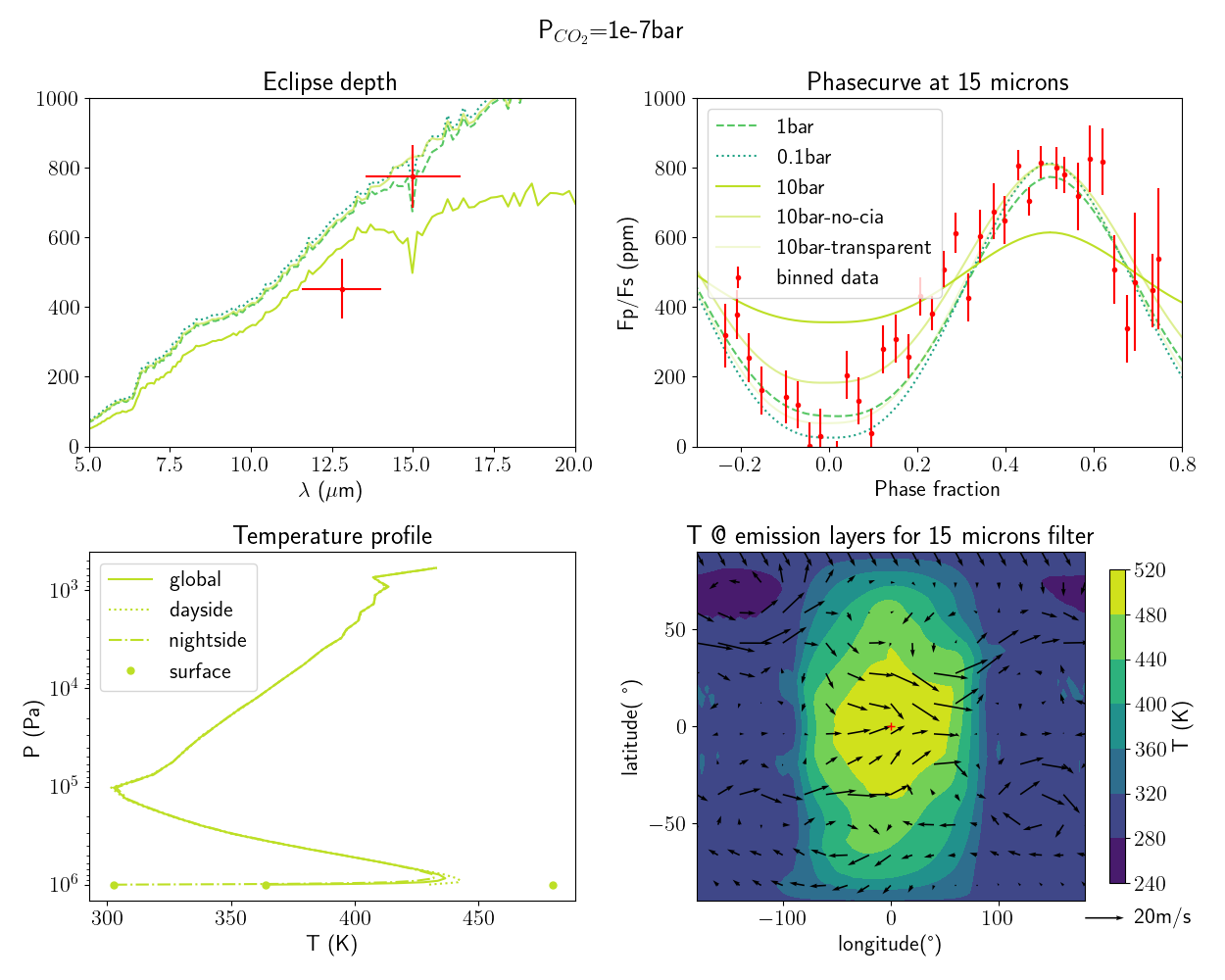}
    \caption{Upper panels: Observables of the (``transparent'' atmospheres. The CO$_2$ partial pressure is 1.e-7~bar, except for the (``transparent'' case where the observables have been calculated for P$_{CO_2}$=1.e-15~bar. For the ``no CIA'' case, the collision-induced absorptions have not been included in the opacity computation. Lower panels: Temperature profile and map for the 10~bar case. The atmosphere is very well mixed, but there is no efficient coupling with the surface, and the surface has a hot dayside and cold nightside.}
    \label{fig:pco2_cste}
\end{figure*}

We studied the influence of P$_\mathrm{tot}$ for a constant partial pressure of CO$_2$ in order to test the thick-transparent atmosphere case. Fig.~\ref{fig:pco2_cste} shows the observables for different tests of atmospheres from the ``transparent'' family, and the temperature distribution for the 10~bar, N$_2$+0.1ppm CO$_2$ atmosphere.
One can see that the surface is well decoupled from the atmosphere, since the temperature profile of the atmosphere is almost the same in the nightside and dayside, although the surface presents strong differences between day and night (lower left panel).

 For a fully transparent atmosphere, the heat redistribution should not be detectable in the phase curve as the emission comes from the surface. Yet, atmospheres composed of real gases can not be completely transparent, and even a small opacity allows for the atmosphere to absorb and re-emit part of the radiation from the surface, where the heat is redistributed and the nightside warmer. Thus, one can see that the 10 bar atmosphere phase curve has a weaker amplitude, with a nightside flux of almost 400 ppm. Part of the absorption is caused by collision-induced absorption of N$_2$, and part of it by the remaining CO$_2$ present in the atmosphere. When we remove these contributions (``transparent'' case in the figure), the phase curve is similar to a thin atmosphere.

\subsection{Pure CO$_2$}\label{sec:app-co2}
\begin{figure*}[h!]
    \centering
    \includegraphics[width=\linewidth]{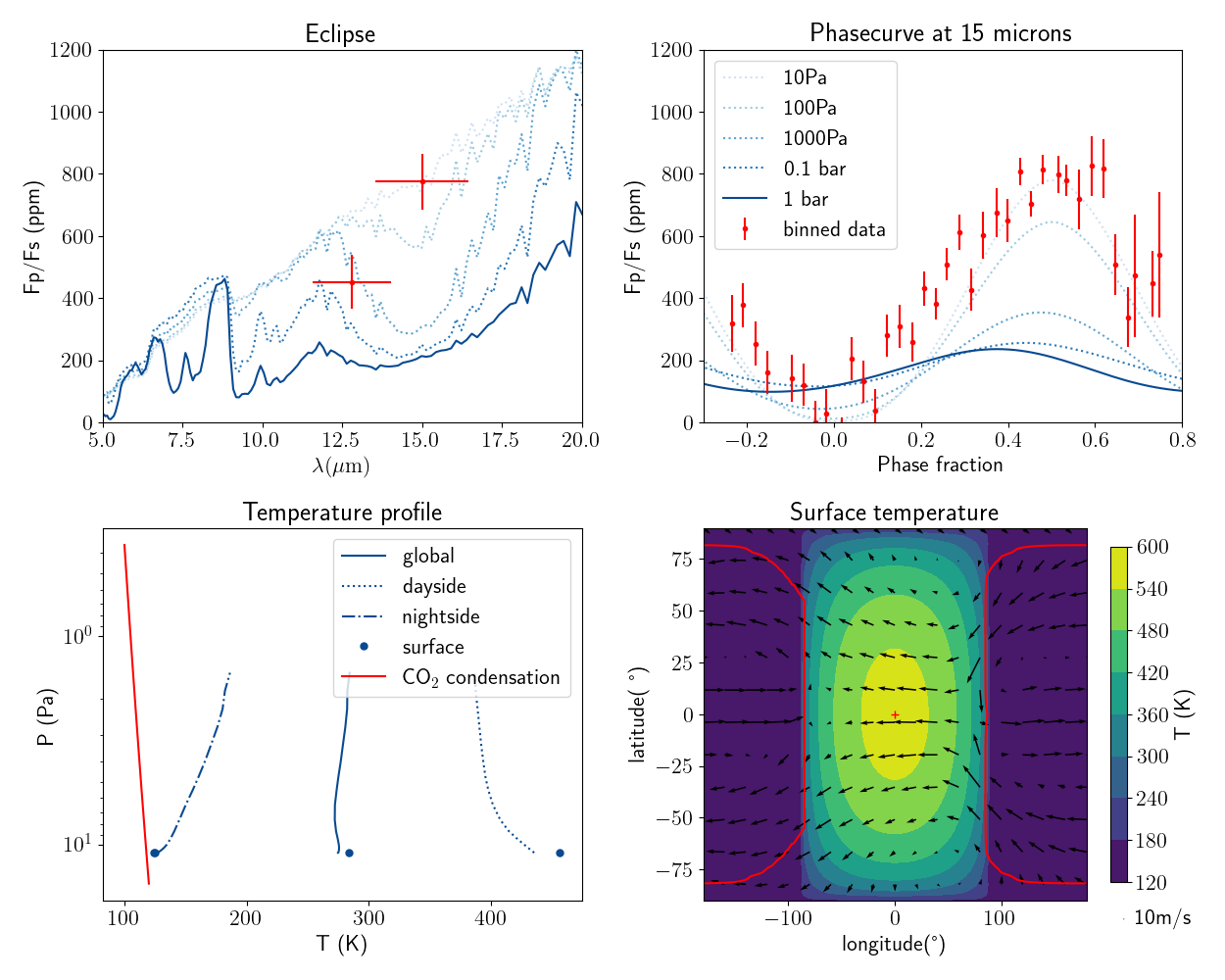}
    \caption{Pure CO$_2$ case. Upper panels: Observables (right: phase curve at 15~$\mu$m, left: eclipse depth). The dotted lines represent non-stable atmospheres; these are snapshots of the atmosphere during its collapse. Lower panels: Physical quantities. Temperature profiles and temperature map for the 10~Pa case (compatible with the observations, but not stable). In the lower right panel, the red contour shows the CO$_2$ ice caps. The red dot is the substellar point. In the lower left panel, the red line corresponds to the CO$_2$ condensation curve.}
    \label{fig:obs_co2_collapse}
\end{figure*}

We show in Fig.~\ref{fig:obs_co2_collapse} the observables and temperature for the case of a collapsing, pure CO$_2$ atmosphere, with the boundary of the CO$_2$ ice cap.

\subsection{Effect of surface albedo}\label{sec:app-albedo}
\begin{figure}[h!]
     \centering
     \begin{subfigure}[b]{0.47\textwidth}
         \centering
         \includegraphics[width=\textwidth]{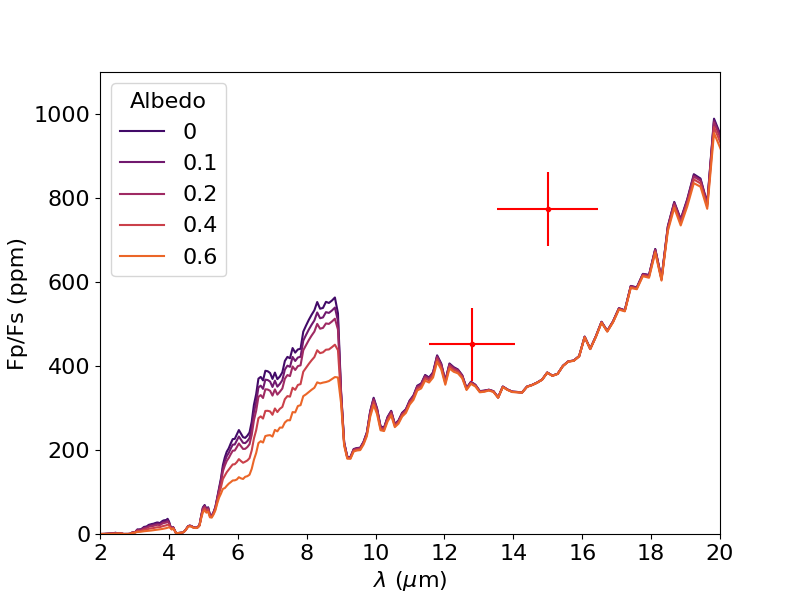}
         \caption*{(a) Pure CO$_2$, 1~bar}
     \end{subfigure}
     \hfill
          \begin{subfigure}[b]{0.47\textwidth}
         \centering
         \includegraphics[width=\textwidth]{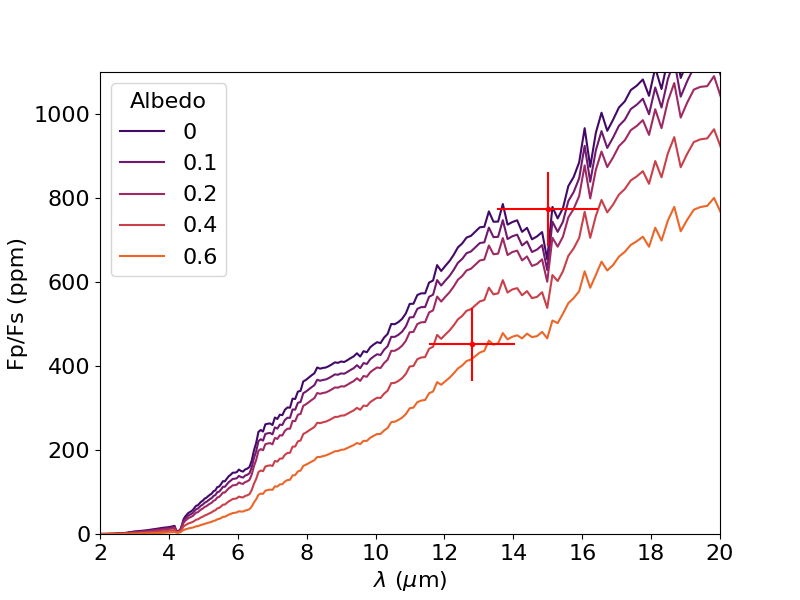}
         \caption*{(b) 1 ppm CO$_2$, 1~bar}
     \end{subfigure}
     \hfill
     \begin{subfigure}[b]{0.47\textwidth}
         \centering
         \includegraphics[width=\textwidth]{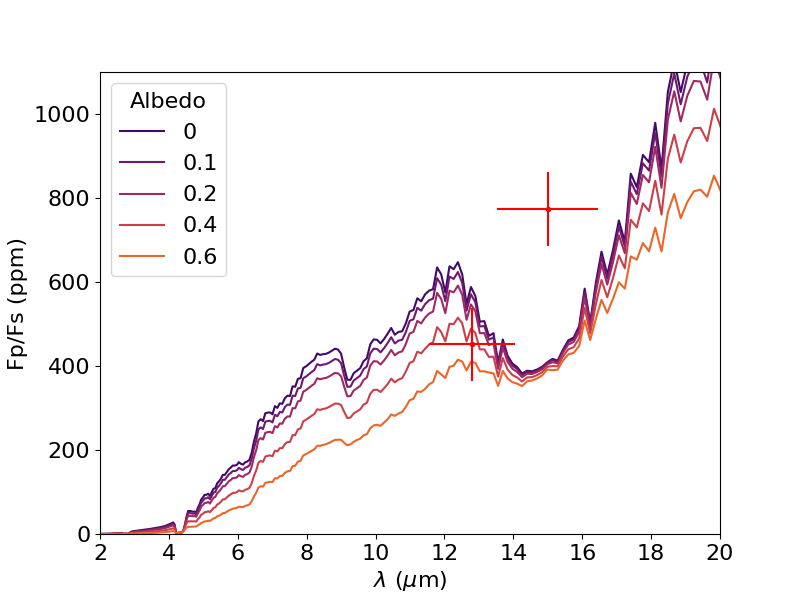}
         \caption*{(c) Pure CO$_2$, 0.01~bar}
     \end{subfigure}
     \hfill
     \begin{subfigure}[b]{0.47\textwidth}
         \centering
         \includegraphics[width=\textwidth]{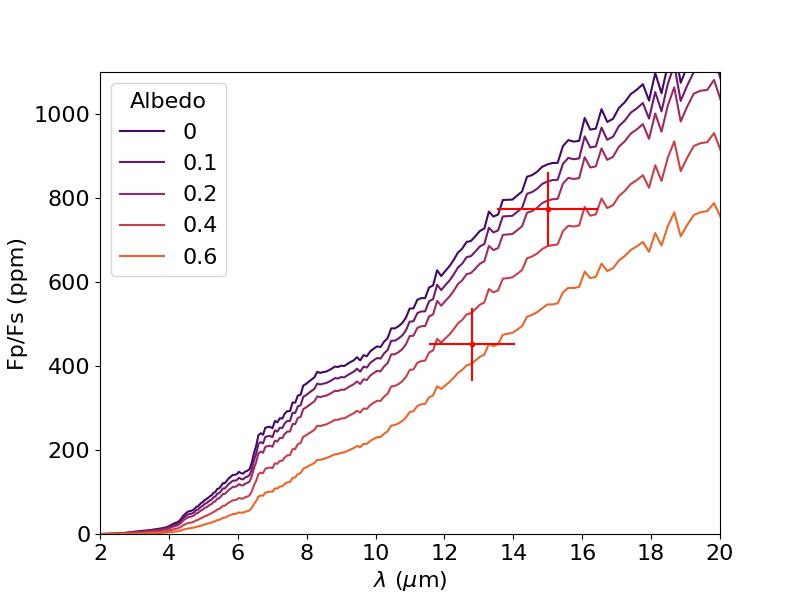}
         \caption*{(d) 1 ppm CO$_2$, 0.01~bar}
     \end{subfigure}
        \caption{Effect of the surface albedo for different cases of N$_2$-CO$_2$ atmospheres (results of 1D simulations). }
        \label{fig:albedo}
\end{figure}

We studied the effect of the surface albedo for several atmospheres from the CO$_2$-N$_2$ grid in Fig.~\ref{fig:albedo}. If the CO$_2$ absorption band is strong (a,c), the atmosphere is optically thick at 15~$\mu$m and the surface albedo does not matter at this wavelength. For cases with little CO$_2$ proportion (b,d), there is no saturation effect and the albedo lowers the whole spectrum equally.
We can then suppose that for low CO$_2$ proportion, increasing the albedo can decrease the eclipse depth for 12.8 and 15~$\mu$m, although for high CO$_2$ proportion, increasing the albedo would decrease the eclipse depth at 12.8~$\mu$m, but not at 15~$\mu$m.

\clearpage

\section{Thick atmosphere with an opacity window}\label{sec:app/thick}

\subsection{Proof of concept}\label{sec:app/thick/concept}
\begin{figure*}[h!]
    \centering
    \includegraphics[width=0.9\linewidth]{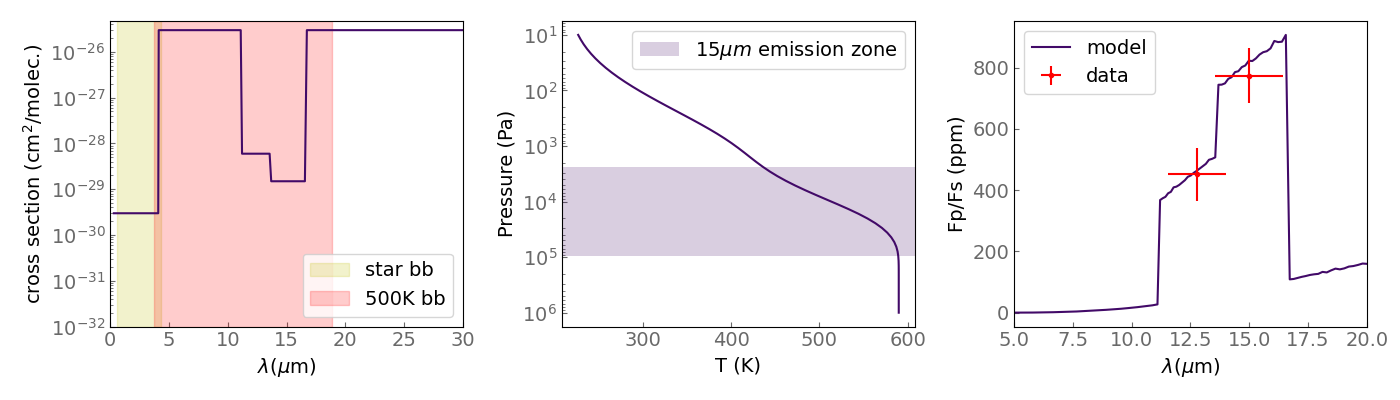}
    \caption{1D study of thick atmosphere with greenhouse effect with an opacity window at the observed wavelengths. Left: Idealized opacity vs wavelength. The yellow window corresponds to the zone of emission of the star, the red window to the planet. Middle: Temperature profile. The window corresponds to the layers of 95\% of the emission. Right: Corresponding eclipse depth.}
    \label{fig:thick_1D_fke_op}
\end{figure*}

Figure~\ref{fig:thick_1D_fke_op} shows a 1D proof-of concept of the family of the thick atmospheres. We designed an unrealistic opacity of the atmosphere meeting our requirements: a high opacity in the range of thermal emission of the planet, a window around the probed wavelengths, and a weaker opacity at the range of emission of the star, to let the stellar flux reach the surface of the planet. We ran the 1D radiative transfer code \texttt{exo\_k} and tuned the values of the opacity to reach the eclipse observation points. Thanks to the absorption window, the main part of the emission at 12.8 and 15~$\mu$m comes from deep in the atmosphere, where the temperature corresponds to the measured brightness temperature of the observations. 

\FloatBarrier

\subsection{Opacities}\label{sec:app/thick/opacity}
\begin{figure*}[h!]
    \centering
    \includegraphics[width=0.75\linewidth]{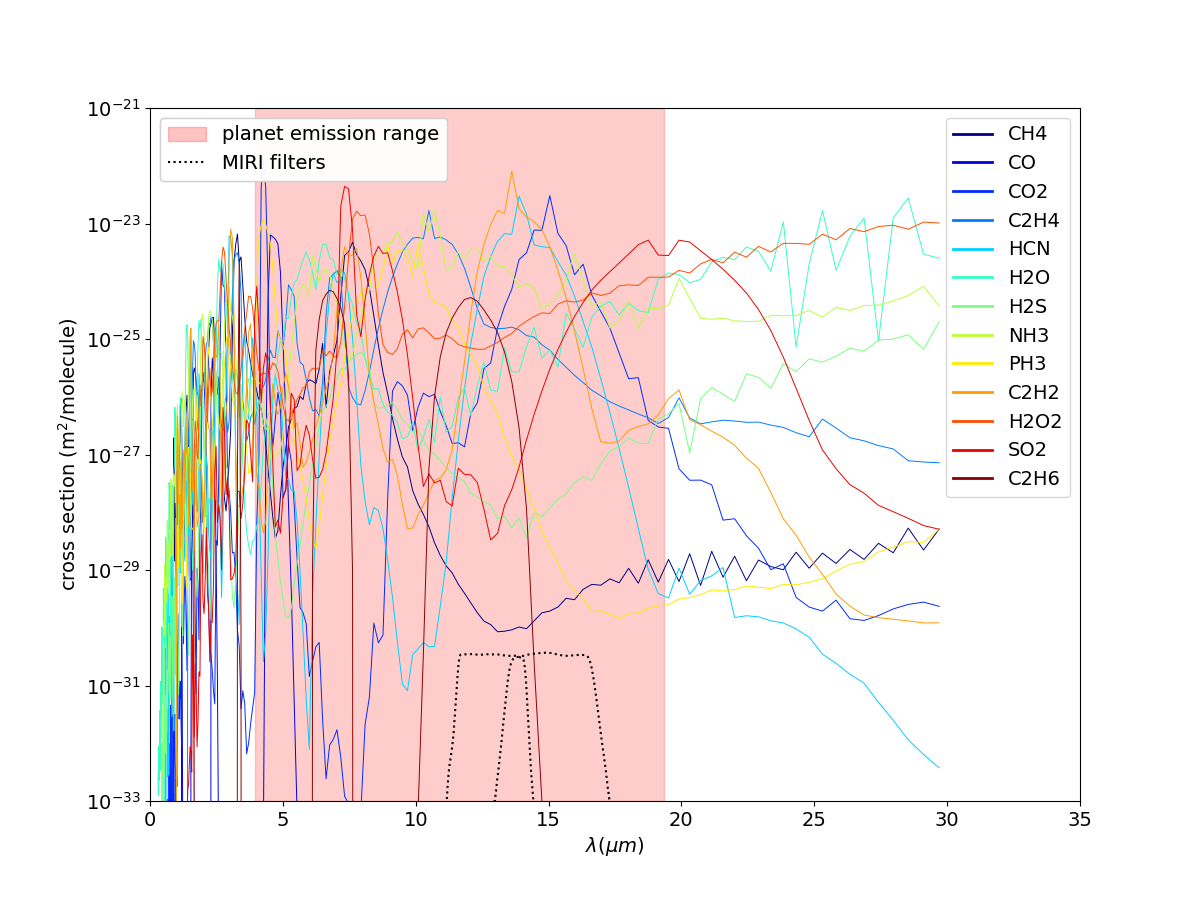}
    \vspace{-0.5em}
    \caption{Opacities of molecules. The dotted lines show the MIRI filters where the molecules must not absorb too much in order to produce the right effect. The planet blackbody emission range is in red.}
    \label{fig:molec_opacities}
\end{figure*}

We show in Fig.~\ref{fig:molec_opacities} the opacities of different molecules. We look for molecules that absorb in most of the planet emission range, but less in the MIRI filters at 12.8 and 15~$\mu$m. C$_2$H$_4$ and C$_2$H$_6$ present these features.

\FloatBarrier

\subsection{3D effect}\label{sec:app/thick/3D}
\begin{figure*}[h!]
    \centering
    \includegraphics[width=0.7\linewidth]{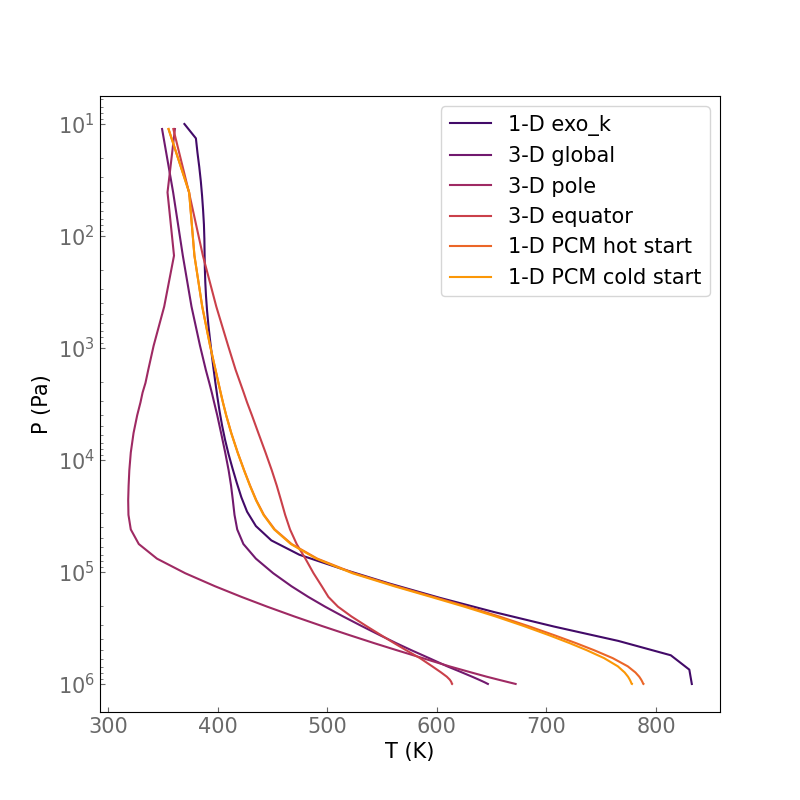}
    \caption{Thermal profiles for different simulations of a 10-bar N$_2$ atmosphere with 20\% CH$_4$ and 0.2\% C$_2$H$4$.}
    \label{fig:thick_comp_profiles}
\end{figure*}

We provide the thermal profiles in 1D and 3D of different simulations in Fig.~\ref{fig:thick_comp_profiles} in order to illustrate the 3D effect resulting in a mean thermal profile colder than 1D computations. In 1D, the deep atmosphere is convective, and the thermal profiles follows an adiabat. In addition to the usual \texttt{exo\_k} simulations, we performed two 1D \texttt{PCM} simulation, one starting from the \texttt{exo\_k} thermal profile (\textit{hot start}) and the other from the 3D mean profile (\textit{cold start}), in order to make sure the difference between the 3D \texttt{Generic PCM} simulations and the 1D simulations comes from 3D effects. The thermal profile of the 1D \texttt{PCM} indeed converges  towards a profile close to the \texttt{exo\_k} profile.\\
In 3D, the deep atmosphere is well mixed horizontally, but the poles are colder than the equator region. This leads to subsidence at the poles, forcing an adiabat that is colder than in the 1D case. This changes the circulation regime, counter-intuitively making the surface temperature colder at the equator than at the poles, and preventing convection at the equator region.

\clearpage
\section{1D study of hazes}\label{sec:app/hazes}
\subsection{Parameter study}\label{sec:app/hazes/parameters}
We conducted a study in 1D with \texttt{exo\_k}. First, we varied different parameters for of the simplified model presented in \ref{sec:methods/gPCM/hazes}. Then, we compared the model with different aerosol optical properties (extinction coefficient, single scattering albedo, and asymmetry factor) from laboratory measurements, and finally we varied some parameters for two selected aerosol types. The parameters of each 1D simulations are summarized in Table \ref{table:app/haze_param_study}.

\begin{table*}[h!]
    \caption{Parameters for the aerosol parameter study.}
    \centering
    \begin{tabular}{lccccccr}
       \hline
       Study & Aerosol type & P$_\mathrm{bot}$ (Pa) & P$_\mathrm{top}$ (Pa) & $f_\mathrm{haze}$ & VMR & SSA & r$_\mathrm{eff}$ (m)\\
       \hline
    a & Simplified & $1 \times 10^2$ & $1.\times 10^0$ & $7.\times 10^{-4}$ & -- & 0 & --\\
     &  Simplified & $1 \times 10^3$ & $1.\times 10^1$ & $7.\times 10^{-4}$ & -- & 0 & --\\
     &  Simplified & $1 \times 10^4$ & $1.\times 10^0$ & $7.\times 10^{-4}$ & -- & 0 & --\\
     &  Simplified & $1 \times 10^4$ & $1.\times 10^1$ & $7.\times 10^{-4}$ & -- & 0 & --\\
     &  Simplified & $1 \times 10^5$ & $1.\times 10^0$ & $7.\times 10^{-4}$ & -- & 0 & --\\
       \hline
    b & Simplified & $1 \times 10^5$ & $1.\times 10^0$ & $1.\times 10^{-5}$ & -- & 0 & --\\
      & Simplified & $1 \times 10^5$ & $1.\times 10^0$ & $1.\times 10^{-4}$ & -- & 0 & --\\
      & Simplified & $1 \times 10^5$ & $1.\times 10^0$ & $1.\times 10^{-2}$ & -- & 0 & --\\
      & Simplified & $1 \times 10^5$ & $1.\times 10^0$ & $7.\times 10^{-4}$ & -- & 0 & --\\
        \hline
    c & Tholins (Khare) & $1 \times 10^3$ & $1 \times 10^0$ & -- & $1 \times 10^{-13}$ & free & $1 \times 10^{-6}$\\
      & Tholins (Drant) & $1 \times 10^3$ & $1 \times 10^0$ & -- & $1 \times 10^{-13}$ & free & $1 \times 10^{-6}$\\
      & H$_2$SO$_4$     & $1 \times 10^3$ & $1 \times 10^0$ & -- & $1 \times 10^{-13}$ & free & $1 \times 10^{-6}$\\
      & Dust            & $1 \times 10^3$ & $1 \times 10^0$ & -- & $1 \times 10^{-18}$ & free & $1 \times 10^{-4}$\\
      & Simplified & $1 \times 10^5$ & $1.\times 10^0$ & $7.\times 10^{-4}$ & -- & 0 & --\\
      \hline
    d & Tholins (Khare) & $1 \times 10^3$ & $1 \times 10^0$ & -- & $1 \times 10^{-13}$ & 0.5 & $1 \times 10^{-6}$\\
      & Tholins (Drant) & $1 \times 10^3$ & $1 \times 10^0$ & -- & $1 \times 10^{-13}$ & 0.5 & $1 \times 10^{-6}$\\
      & H$_2$SO$_4$     & $1 \times 10^3$ & $1 \times 10^0$ & -- & $1 \times 10^{-13}$ & 0.5 & $1 \times 10^{-6}$\\
      & Dust            & $1 \times 10^3$ & $1 \times 10^0$ & -- & $1 \times 10^{-18}$ & 0.5 & $1 \times 10^{-4}$\\
      & Simplified & $1 \times 10^5$ & $1.\times 10^0$ & $7.\times 10^{-4}$ & -- & 0 & --\\
        \hline
    e & Tholins (Khare) & $1 \times 10^3$ & $1 \times 10^0$ & -- & $1 \times 10^{-13}$ & free & $1 \times 10^{-8}$\\
      & Tholins (Khare) & $1 \times 10^3$ & $1 \times 10^0$ & -- & $1 \times 10^{-13}$ & free & $1 \times 10^{-7}$\\
      & Tholins (Khare) & $1 \times 10^3$ & $1 \times 10^0$ & -- & $1 \times 10^{-13}$ & free & $1 \times 10^{-6}$\\
      & Dust            & $1 \times 10^3$ & $1 \times 10^0$ & -- & $1 \times 10^{-18}$ & free & $1 \times 10^{-8}$\\
      & Dust            & $1 \times 10^3$ & $1 \times 10^0$ & -- & $1 \times 10^{-18}$ & free & $1 \times 10^{-6}$\\
      & Dust            & $1 \times 10^3$ & $1 \times 10^0$ & -- & $1 \times 10^{-18}$ & free & $1 \times 10^{-4}$\\
      \hline
    f & Tholins (Khare) & $1 \times 10^3$ & $1 \times 10^0$ & -- & $1 \times 10^{-18}$ & free & $1 \times 10^{-6}$\\
      & Tholins (Khare) & $1 \times 10^3$ & $1 \times 10^0$ & -- & $1 \times 10^{-15}$ & free & $1 \times 10^{-6}$\\
      & Tholins (Khare) & $1 \times 10^3$ & $1 \times 10^0$ & -- & $1 \times 10^{-12}$ & free & $1 \times 10^{-6}$\\
      & Dust            & $1 \times 10^3$ & $1 \times 10^0$ & -- & $1 \times 10^{-21}$ & free & $1 \times 10^{-4}$\\
      & Dust            & $1 \times 10^3$ & $1 \times 10^0$ & -- & $1 \times 10^{-18}$ & free & $1 \times 10^{-4}$\\
      & Dust            & $1 \times 10^3$ & $1 \times 10^0$ & -- & $1 \times 10^{-15}$ & free & $1 \times 10^{-4}$\\
    \end{tabular}
        \label{table:app/haze_param_study}
    \end{table*}

\begin{figure*}
     \centering
     \begin{subfigure}[b]{0.49\textwidth}
         \centering
         \includegraphics[width=\textwidth]{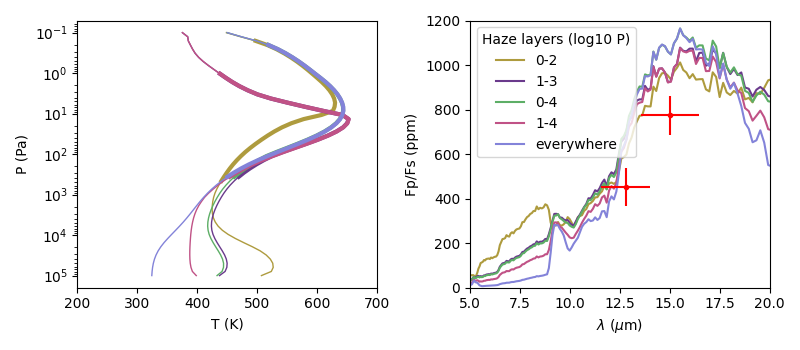}
         \caption{Vertical profile of hazes. The haze VMR is constant between two layers of pressure, and set to 0 outside this band.}
         \label{fig:haze_1}
     \end{subfigure}
     \hfill
     \begin{subfigure}[b]{0.49\textwidth}
         \centering
         \includegraphics[width=\textwidth]{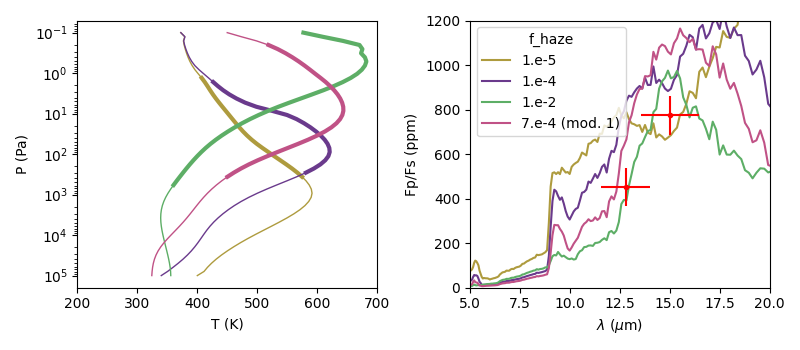}
         \caption{Haze factor $f_\mathrm{haze}$. For this case, the haze vertical profile is a constant.}
         \label{fig:haze_2}
     \end{subfigure}
     \hfill
     \begin{subfigure}[b]{0.49\textwidth}
         \centering
         \includegraphics[width=\textwidth]{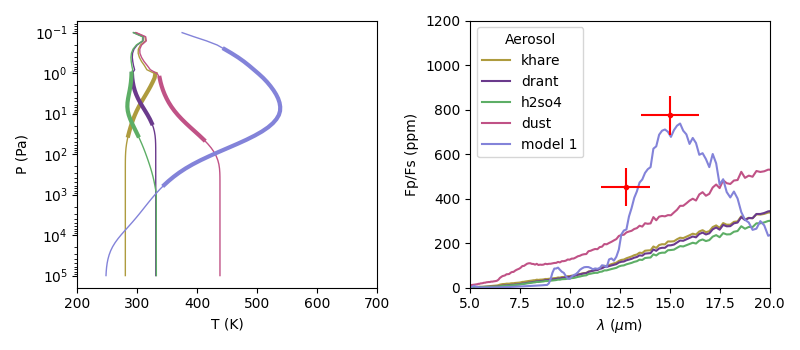}
         \caption{Aerosol optical properties. For all cases, the aerosol VMR is constant between P=1000 and 1~Pa.}
         \label{fig:haze_3}
     \end{subfigure}
     \hfill
          \begin{subfigure}[b]{0.49\textwidth}
         \centering
         \includegraphics[width=\textwidth]{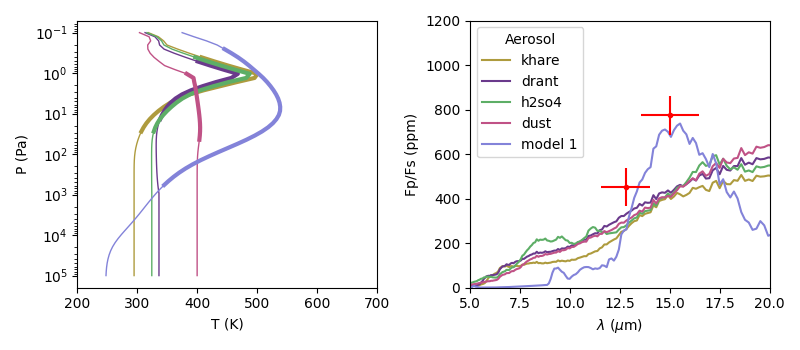}
         \caption{Aerosol optical properties, with single-scattering albedo fixed to 0.5. The parameters are the same as panel (c), but with a single scattering albedo set to 0.5, independent of wavelength.}
         \label{fig:haze_4}

     \end{subfigure}
     \hfill
     \begin{subfigure}[b]{0.49\textwidth}
         \centering
         \includegraphics[width=\textwidth]{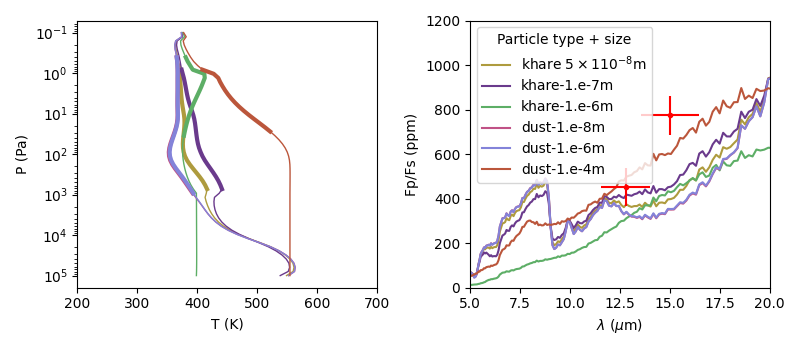}
         \caption{Particle size. Comparison of different radii of particle for two aerosol types: dust, and tholins.}
         \label{fig:haze_5}
     \end{subfigure}
     \hfill
     \begin{subfigure}[b]{0.49\textwidth}
         \centering
         \includegraphics[width=\textwidth]{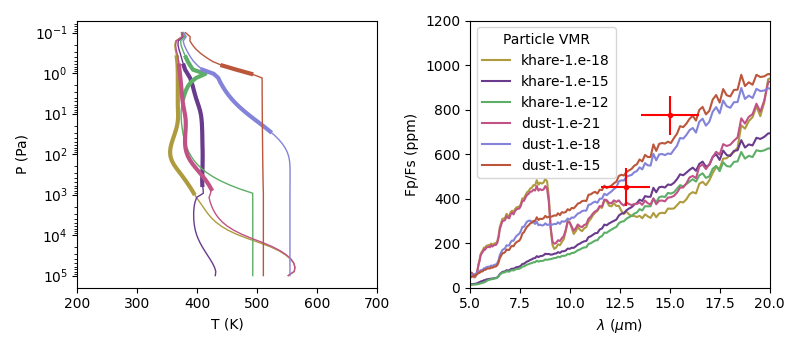}
         \caption{Particle volume mixing ratio. Comparison of different mixing ratios for dust and tholins.}
        \label{fig:haze_6}

     \end{subfigure}
        \caption{Temperature profile and eclipse depth for different parameterizations of aerosols. Each panel illustrates the influence of a parameter.}
        \label{fig:1D_study_hazes}
\end{figure*}

\begin{enumerate}
    \item Impact of the vertical distribution profile of the hazes (Fig.~\ref{fig:haze_1}). We find that the most important is that the hazes are present at the top of the atmosphere. If the upper layer of the atmosphere is hazeless, the CO$_2$ present in this layer is seen in absorption. The presence or absence of hazes in the deeper atmosphere is not important.
    \item Impact of the factor $f_{haze}$ (Fig.~\ref{fig:haze_2}). This factor corresponds to the quantity of hazes present in the atmosphere. We find, like \cite{Ducrot:2024}, that 7.e-4 is optimal, as a lower quantity of hazes results in a thermal inversion deeper in the atmosphere, and the CO$_2$ band probes in the upper atmosphere. Thus, we confirm that the case we present is fine-tuned.
    \item Comparison with real aerosol optical properties (Fig.~\ref{fig:haze_3}). For all the different aerosols that we could test, we do not get a CO$_2$ emission band. Martian dust results in a thermal inversion allowing for a high eclipse depth.
    \item Impact of the single-scattering albedo  (Fig.~\ref{fig:haze_4}). With a single scattering albedo fixed to 0.5, independent of the wavelength, we can reproduce some features of our simplified model.
    \item Impact of the size of the aerosols  (Fig.~\ref{fig:haze_5}). We find, for two different types of aerosols (tholins and dust), that the largest aerosols give inversions upper in the atmosphere, and thus highest eclipse depth.
    \item Impact of the volume mixing ratio  (Fig.~\ref{fig:haze_6}). We varied the volume mixing ratio (VMR) for the same two types of aerosols. For the dust case, the highest VMR leads to the highest eclipse depth, although there is a saturation at $\sim 1 \times 10^{-15}$. For tholins, for VMR $\sim 1 \times 10^{-12}$ there is an inversion, but it is not strong enough to give a high eclipse depth.

\end{enumerate}

\begin{figure}
    \centering
    \includegraphics[width=0.5\linewidth]{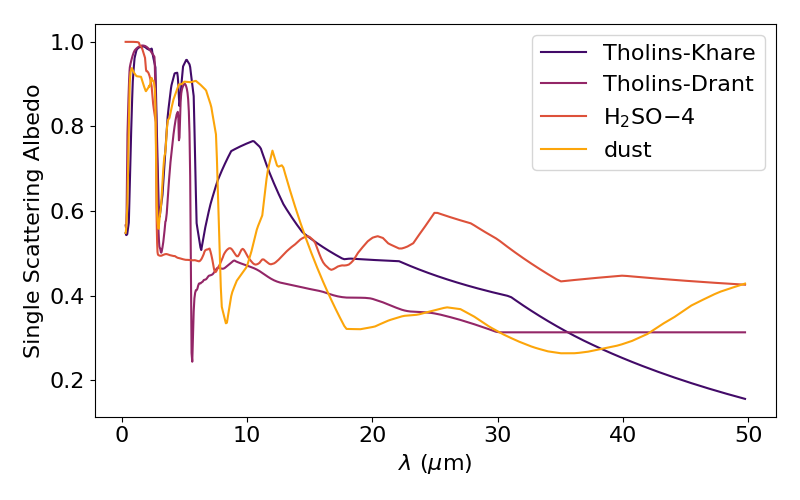}
    \caption{Single scattering albedo vs wavelength for different aerosol types. The radii of aerosols are $10\times 10^{-6}$ for tholins and H$_2$SO$_4$ and $10\times 10^{-4}$ for dust.}
    \label{fig:haze_ssa}
\end{figure}

With these results, we identify that the realistic single scattering albedo of the aerosols, that is very high in the shortest wavelengths (Fig.~\ref{fig:haze_ssa}), is responsible for the absence of CO$_2$ seen in emission by reflecting a very large part of the stellar emission, prohibiting a strong heating of the upper atmosphere.

In all our cases, either there is no strong thermal inversion in the upper layers and CO$_2$ is seen in absorption, or the hazes seem to be optically thick at all wavelength, resulting in an isotherm in the deep atmosphere instead of a thermal inversion, and a thermal emission that reproduces a blackbody radiation with the temperature of the upper atmosphere.

In conclusion, only dust type aerosols are able to reproduce a strong eclipse depth, and no aerosol reproduces the CO$_2$ emission band that is seen with the simplistic haze model.


\FloatBarrier

\subsection{Convergence issues}\label{sec:app/hazes/convergence}
\begin{figure}[h!]
    \centering
    \includegraphics[width=0.4\linewidth]{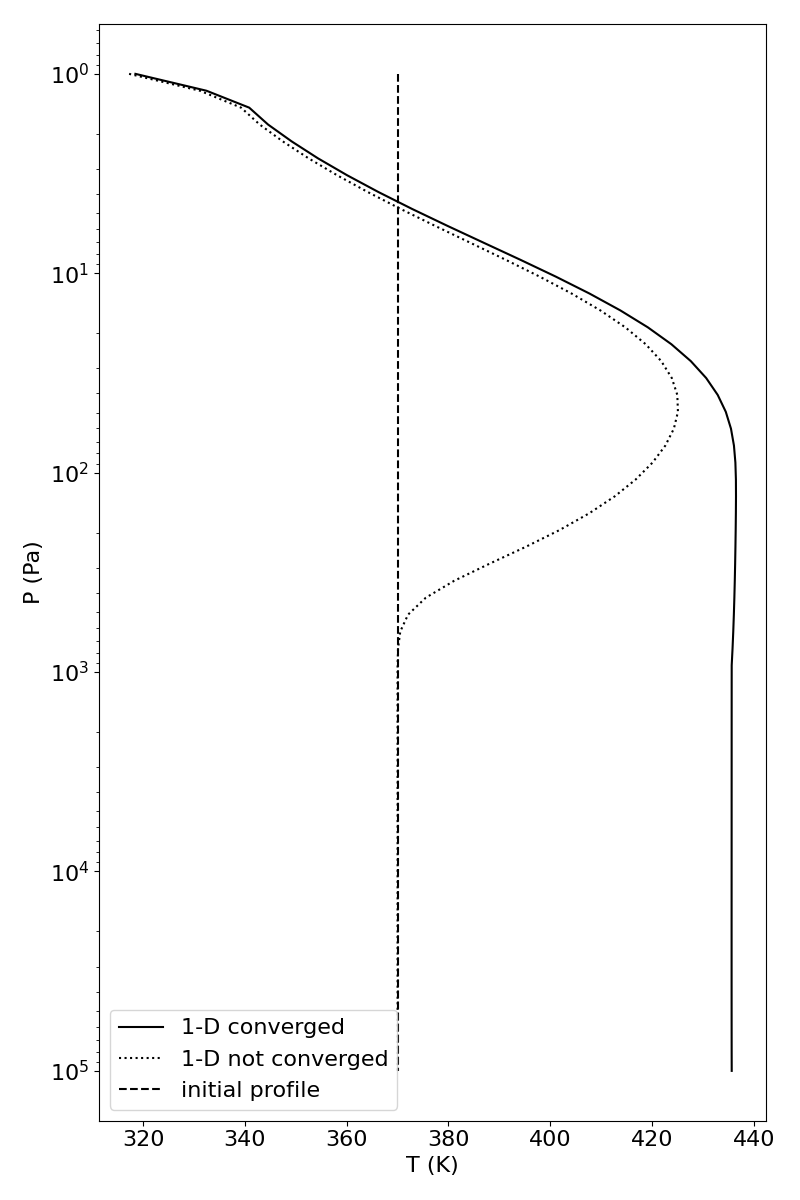}
    \caption{Thermal 1D profile for the dust case, converged vs not converged. The initial profile is a 370~K isotherm.}
    \label{fig:cvgence_aerosol}
\end{figure}

Some of these simulations contain a high amount of aerosols, and only a very small part of the radiation goes through the atmosphere. Therefore, the deep atmosphere is getting almost no flux, and its convergence is extremely long. The time step of the simulations is necessarily short, because of the small radiative timescale at the high atmosphere, whereas in the deep atmosphere the evolution timescale is way longer than the time step. \texttt{exo\_k} comes with several acceleration routines, making it possible to converge this type of atmosphere. However, the \texttt{Generic PCM} does not, and our 3D simulation of the dust scenario is not converged in the deep atmosphere. This does not impact our conclusion on observables as the thermal emission comes from the upper parts of the atmosphere, that are converged. Fig.~\ref{fig:cvgence_aerosol} shows that the 1D \texttt{exo\_k} profile is similar, when not converged, to the 3D profile.

\FloatBarrier

\section{Transit spectra}\label{sec:app/transit}
We computed transit spectra for representatives of the families that match the eclipse depths and phase curve: N$_2$-CO$_2$ atmospheres (here, 1 bar with 1ppm of CO$_2$), and atmospheres with thermal inversion (here, the hazes from the simplified hazes Model 1). The hazes distinguish from the bare rock and N$_2$-CO$_2$ case by a slope in the shortest wavelengths. However, this feature is of the order of dozens of ppm which is for now too dim to be observed. Nevertheless, these features could be of importance for future observations of the system, especially when TRAPPIST-1 b is used as a probe to correct for stellar contamination.

\begin{figure}[h!]
    \centering
    \includegraphics[width=0.5\linewidth]{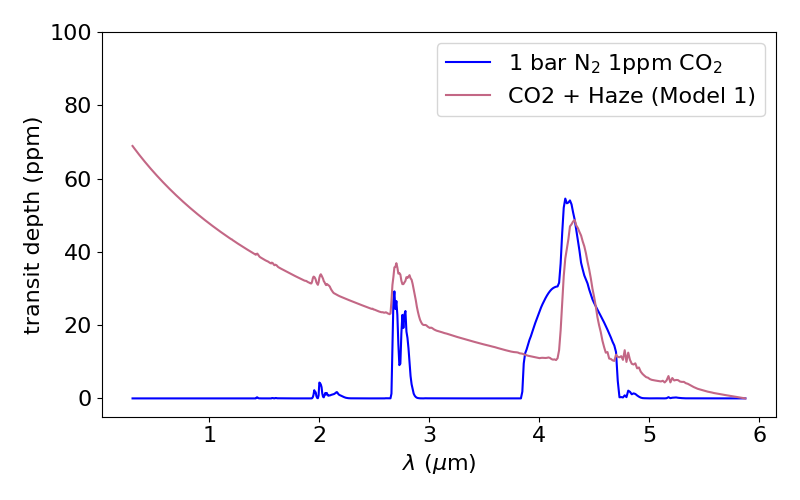}
    \caption{Transit spectra computed for scenarios that match all the available emission observations. The transit spectrum has been offset to have a minimum value of 0.}
    \label{fig:transmission}
\end{figure}

\end{appendix}

\end{document}